\documentclass[a4paper,11pt]{article}
\pdfoutput=1 
\usepackage{jheppub} 
                     
\usepackage{booktabs,makecell}
\usepackage[font=small]{caption}
\usepackage{framed}
\usepackage{hyperref}
\usepackage[export]{adjustbox}
\usepackage{amsmath}
\DeclareMathOperator{\Tr}{Tr}
\usepackage{physics}
\usepackage{subcaption}

\def\CN{\mathcal{N}}

\title{Large $N$ Universality of 4d $\CN=1$ Superconformal Index and AdS Black Holes}

\author[a]{Sunjin Choi,}
\author[b]{Seunggyu Kim,}
\author[b]{and Jaewon Song}

\affiliation[a]{School of Physics, Korea Institute for Advanced Study \\85 Hoegi-ro, Dongdaemun-gu, Seoul 02455, Republic of Korea}
\affiliation[b]{Department of Physics, Korea Advanced Institute of Science and Technology\\291 Daehak-ro, Yuseong-gu, Daejeon 34141, Republic of Korea}

\emailAdd{sunjinchoi@kias.re.kr}
\emailAdd{sgkim01@kaist.ac.kr}
\emailAdd{jaewon.song@kaist.ac.kr}

\preprint{KIAS-P23034}

\abstract{
We study the large $N$ limit of the matrix models associated with the superconformal indices of four-dimensional $\CN=1$ superconformal field theories. We find that for a large class of $\CN=1$ superconformal gauge theories, the superconformal indices in the large $N$ limit of such theories are dominated by the  `parallelogram' saddle, providing $O(N^2)$ free energy for the generic value of chemical potentials. This saddle corresponds to BPS black holes in AdS$_5$ whenever a holographic dual description is available. Our saddle applies to a large class of gauge theories, including ADE quiver gauge theories, and the theories with rank-2 tensor matters. Our analysis works for most $\CN=1$ superconformal gauge theories that admit a suitable large $N$ limit while keeping the flavor symmetry fixed. 
We also find `multi-cut' saddle points, which correspond to the orbifolded Euclidean black holes in AdS$_5$. 
}

\begin{document} 
\maketitle

\section{Introduction}
One of the crucial implications of the AdS/CFT correspondence \cite{Maldacena:1997re, Gubser:1998bc, Witten:1998qj} is that the boundary conformal field theories (CFTs) should exhibit a universal thermodynamic phase, in which the bulk black hole geometry emerges. This universal property of (holographic) CFTs is independent of the details of the theory \cite{Hawking:1982dh, Witten:1998zw}. One ideal setting to study such a phenomenon is to focus on the BPS sector, which has been an extremely useful setting, as was demonstrated in the microscopic derivation of the Bekenstein-Hawking entropy by Strominger-Vafa \cite{Strominger:1996sh}. We would like to take a similar strategy to understand bulk geometry from the BPS sector of the boundary. In the BPS sector, one can exactly compute the BPS observables of superconformal field theories (SCFTs); hence, quantitative analysis can be carried out.

Relatedly, another pressing question in AdS/CFT is to find a precise condition for the CFT in which the holographic bulk description in terms of a weakly-coupled Einstein gravity emerges. For the case of two-dimensional CFTs, such conditions have been discussed in \cite{Hartman:2014oaa} utilizing the modular invariance of the thermal partition function. Assuming the existence of Hawking-Page-like phase transition between the graviton gas phase and the large AdS black hole phase, the authors of \cite{Hartman:2014oaa} were able to derive the Bekenstein-Hawking entropy formula, which has the same form as the Cardy formula. However, the crucial difference here is that the Cardy formula has a very different regime of validity (finite central charge, very large scaling dimension $\Delta \gg c$) from the large AdS black hole (large central charge $c \gg 1$, scaling dimension of order $\Delta > c/12 $). A similar analysis was also carried out for SCFTs in 2d using the elliptic genus \cite{Benjamin:2015hsa, Benjamin:2015vkc}. 

We would like to tackle similar questions in the AdS$_5$/CFT$_4$ context. Which 4d SCFTs are holographically dual to a weakly-coupled Einstein gravity? When do we have emergent black hole geometry? More modestly, we aim to find SCFTs with a phase potentially described by the large BPS black holes in AdS. To this end, we shall study the large $N$ behavior of the superconformal index \cite{Romelsberger:2005eg,Kinney:2005ej} for the 4d $\mathcal{N}=1$ SCFTs, which are realized from (the IR fixed point of) $\mathcal{N}=1$ supersymmetric gauge theories.
The superconformal index enumerates the $(-1)^F$-graded degeneracy of certain BPS states in the radially quantized SCFT on $S^3$. 
One of the original motivations of the index is to account for the Bekenstein-Hawking entropy of the BPS black holes in AdS$_5$ \cite{Gutowski:2004ez,Gutowski:2004yv,Chong:2005hr,Kunduri:2006ek}. 
This goal was realized a decade later by a series of works utilizing the complex saddles of the large $N$ index from various viewpoints starting from \cite{Cabo-Bizet:2018ehj, Choi:2018hmj, Benini:2018ywd}. Subsequently, various developments have been made using different methods, such as Cardy-like limits\footnote{Pioneering work on the Cardy-like limit of the index were done in \cite{DiPietro:2014bca, ArabiArdehali:2019tdm}.} (also in finite $N$) \cite{Choi:2018hmj, Choi:2018vbz, Honda:2019cio, ArabiArdehali:2019tdm, Kim:2019yrz, Cabo-Bizet:2019osg, Amariti:2019mgp, Goldstein:2020yvj, Jejjala:2021hlt, ArabiArdehali:2021nsx, Ardehali:2021irq,Cabo-Bizet:2021plf, Mamroud:2022msu}, and Bethe-Ansatz type formula \cite{Closset:2017bse, Benini:2018mlo} for the large $N$ index \cite{Benini:2018ywd, Lanir:2019abx, GonzalezLezcano:2019nca, Benini:2020gjh} and particular extension of the index integral \cite{Cabo-Bizet:2019eaf, Cabo-Bizet:2020nkr}.

The main result of this paper is to show that the matrix model of the index for this large class of gauge theories admits a universal large $N$ saddle point, called the `parallelogram ansatz,' which accounts for the $O(N^2)$ free energy necessary to account for the black holes. It was first discovered for the $\mathcal{N}=4$ $U(N)$ supersymmetric Yang-Mills theory (SYM) in \cite{Choi:2021rxi} that works \emph{beyond the particular limit} (such as one parameter or certain ratio) of the fugacities in the previous studies, and also showed that it is indeed a \emph{genuine saddle point} of the index integral, without any ad hoc assumption. This large $N$ solution is independent of the details of the theory and only depends on the angular momentum fugacities on $S^3$, thus being universal. We demonstrate how it can be generalized to $\mathcal{N}=1$ gauge theories, which allows generic holographic theories to account for the universal BPS black holes in AdS$_5$ \cite{Gutowski:2004ez,Gutowski:2004yv,Chong:2005hr,Kunduri:2006ek}.

We find that such saddle point exists for a large class of gauge theories admitting a suitable large $N$ limit. In particular, it works for a gauge theory with a finite number of matter fields in arbitrary rank-one or rank-two representations under the gauge group. It works for all simple large $N$ gauge theories with a fixed flavor symmetry, which is classified in \cite{Agarwal:2020pol}. The saddle also works for the quiver gauge theories with $SU(N)$, $SO(N)$, and $Sp(N)$ gauge nodes. The prime examples are the Klebanov-Witten theory \cite{Klebanov:1998hh}, toric models such as the $Y^{p,q}$-theories \cite{Martelli:2004wu,Benvenuti:2004dy}, and, of course, the $\mathcal{N}=4$ super-Yang-Mills theories with $SU(N)$, $SO(N)$, or $Sp(N)$ gauge group. 
The `parallelogram saddle' works for most of the known holographic Lagrangian gauge theories, and accounts for the $O(N^2)$ free energy of the AdS black holes.  
The list of theories that do not admit our saddle point includes supersymmetric quantum chromodynamics (SQCD) in the conformal window, class $\mathcal{S}$ theories with full punctures \cite{Gaiotto:2009we}, and the theories with a `dense spectrum' \cite{Agarwal:2019crm, Agarwal:2020pol}. This does not rule out the possibility of another kind of saddle responsible for the black hole phase. 

The superconformal index we study is defined as \cite{Kim:2019yrz, Cabo-Bizet:2020nkr, Cassani:2021fyv}
\begin{align}
    \mathcal{I} (\sigma,\tau) = \Tr \left[ e^{-\pi i R}e^{2\pi i \sigma \left(J_1+\frac{R}{2}\right)} e^{2\pi i \tau \left(J_2+\frac{R}{2}\right)} \right]\ ,
\end{align}
where the trace is taken over the Hilbert space of radially quantized SCFT on $S^3$. Here, $J_1$, $J_2$ are the Cartan charges of $SO(4)$ rotation symmetry, and $R$ is the superconformal $U(1)_R$ $R$-charge. We can also include the fugacities for the flavor symmetries. Notice that we have inserted $e^{-\pi i R}$ instead of the usual $(-1)^F$. This version of the index turns out to exhibit a deconfinement phase in the Cardy-like limit $\sigma, \tau \to 0$, whereas the old version does not \cite{Kim:2019yrz, Cabo-Bizet:2019osg}. The former is related to the latter by going to the `second sheet' in the chemical potential space \cite{Cassani:2021fyv}. It was shown that in the Cardy-like limit $\sigma, \tau \to 0$, the index takes the universal expression \cite{Choi:2018hmj, Kim:2019yrz, Cabo-Bizet:2019osg}
\begin{align} \label{eq:Cardy}
    \log\mathcal{I}(\sigma,\tau)\sim -\,(5a-3c) \; \frac{2i\pi\delta^3}{27\sigma\tau} - (a-c)\frac{4 \pi i \delta}{ 3 \sigma \tau} \ , 
\end{align}
which is completely determined by the central charges $a$ and $c$. Here, $\delta \equiv \frac{\sigma+\tau-1}{2}$. This expression holds without taking any large $N$ limit, which serves as an analog of Cardy's formula in two-dimensions \cite{Cardy:1986ie}. The Cardy-like formula reproduces the Bekenstein-Hawking entropy of large BPS black holes in AdS$_5$ \cite{Gutowski:2004ez,Gutowski:2004yv,Chong:2005hr,Kunduri:2006ek} whenever the holographic dual of SCFT is known. However, we know that the black hole threshold ($\Delta \sim c$) is much lower than that of the Cardy limit ($\Delta \gg c$). Therefore, this formula does not a priori apply to the black holes near the threshold. 

In this paper, using the parallelogram ansatz for the index matrix model, we obtain a universal large $N$ formula for the index of 4d $\CN=1$ SCFTs, which is given by
\begin{align}
    \log\mathcal{I}(\sigma,\tau)\sim -\,(5a-3c) \; \frac{2i\pi\delta^3}{27\sigma\tau}\ ,
\end{align}
with $\delta = \frac{\sigma+\tau-1}{2}$. Here, $a\sim c = O(N^2)$, and we only kept the terms of order $O(N^2)$, where $N$ is the rank of the gauge group. Notice that this formula agrees with the Cardy-like formula \eqref{eq:Cardy} without taking the Cardy limit. This expression gives the dominant contribution for the free-energy above the black hole threshold, which is far below the Cardy limit. For holographic theories, the above formula perfectly captures the Bekenstein-Hawking entropy of the universal BPS black holes in AdS$_5$. If we turn on the flavor chemical potentials, the above formula is generalized as
\begin{align}
    \log\mathcal{I} (\Delta,\sigma,\tau)\sim -\,\frac{i\pi}{24\sigma\tau}\sum_{I,J,K}\Tr( Q^IQ^JQ^K)\,\Delta_I\Delta_J\Delta_K\ ,
\end{align}
where $2Q^I$'s are various candidate $R$-charges made by the linear combinations of true superconformal $R$-charge and flavor charges, and $\Delta_I$'s are the corresponding chemical potentials. (For a precise definition, see Section \ref{section1}.) One can see that the ’t Hooft anomaly coefficients completely govern the large $N$ behavior of the index. The above formula precisely matches the entropy function of the BPS black holes in AdS$_5 \times \textrm{SE}_5$ \cite{Hosseini:2017mds, Hosseini:2018dob, Lanir:2019abx, Amariti:2019mgp, Benini:2020gjh}, thus accounting for their microstates.

In addition, generalizing the parallelogram saddle point, we find the multi-cut saddle points, where the eigenvalues of the large $N$ matrix model are clustered to form a finite number of parallelograms \cite{Choi:2021rxi}. For holographic theories, these large $N$ saddles are dual to the orbifolds of the Euclidean BPS black hole solutions in AdS$_5$ \cite{Aharony:2021zkr}. These extra saddles played a crucial role in resolving a version of the `information paradox' of the index analog of the spectral form factor (SFF) \cite{Choi:2022asl}. These saddles are subdominant contributions to the index version of SFF at an `early time,' but it becomes dominant at a `late time.' This is in sharp contrast with the situation in 2d JT gravity, where the spacetime wormholes played a crucial role in recovering the information \cite{Saad:2018bqo, Saad:2019lba}.

The rest of this paper is organized as follows. In Section \ref{section1}, we study the large $N$ limit of the matrix model of the superconformal index for a large class of 4d $\CN=1$ supersymmetric gauge theories, which flow to interacting SCFTs at IR. We show that the `parallelogram ansatz' is the saddle point for a large class of gauge theories. In Section \ref{sec:examples}, we apply the large $N$ analysis in Section \ref{section1} to several concrete holographic models and reproduce the corresponding entropy functions. In Section \ref{sec:discussion}, we conclude with remarks and future directions.


\section{Large $N$ limit of the superconformal index}
\label{section1}

\subsection{Large $N$ saddle point equation from $SL(3, \mathbb{Z})$}\label{sec:sl3z}
In this subsection, we study the large $N$ behavior of the superconformal index of 4d $\CN=1$ gauge theory and obtain the large $N$ saddle point equation, by making use of its $SL(3,\mathbb{Z})$ modular properties. The superconformal index of 4d $\mathcal{N}=1$ SCFT is defined as \cite{Romelsberger:2005eg,Kinney:2005ej}
\begin{equation}\label{eq:tr-index1}
    \mathcal{I}(\delta,\delta^I,\sigma,\tau)=\Tr \left[ e^{2\pi i\sigma J_1}e^{2\pi i\tau J_2} e^{2\pi i \delta R }\prod_{I=1}^{d-1}e^{2\pi i\delta^I f_I} \right]\ ,
\end{equation}
with a constraint \cite{Choi:2018hmj,Kim:2019yrz}
\begin{equation}
    2\delta-\sigma-\tau=-1\ ,
\end{equation}
where the trace is taken over the Hilbert space of radially quantized CFT on $S^3$. Here, $J_1$, $J_2$ are the Cartan charges of $SO(4)$ rotation symmetry, $R$ is the superconformal $U(1)_R$ $R$-charge, and $f_{I=1,2,\cdots,\,d-1}$ are the Cartan charges of the flavor symmetry. Choosing one specific supercharge $\mathcal{Q}$ carrying $(J_1,J_2,R) = (-\frac{1}{2},-\frac{1}{2},+1)$, we find
\begin{align}
    \begin{split}
        e^{2\pi i(\sigma J_1+\tau J_2+\delta R+\sum_{I=1}^{d-1}\delta^If_I)}\mathcal{Q}&=e^{2\pi i\left(\frac{-\sigma-\tau+2\delta}{2}\right)}\mathcal{Q}e^{2\pi i(\sigma J_1+\tau J_2+\delta R+\sum_{I=1}^{d-1}\delta^If_I)}\\
        &=-\, \mathcal{Q}e^{2\pi i(\sigma J_1+\tau J_2+\delta R+\sum_{I=1}^{d-1}\delta^If_I)}\ .
    \end{split}
\end{align}
Note that $\mathcal{Q}$ is not charged under the flavor symmetry. Hence, even though the trace formula \eqref{eq:tr-index1} is not defined using the usual $(-1)^F$ insertion, where $F$ is the fermion number operator, it is indeed an index \cite{Choi:2018hmj, Kim:2019yrz, Cassani:2021fyv}. Namely, it receives contribution only from the $\frac{1}{4}$-BPS states satisfying $\{\mathcal{Q},\mathcal{Q}^\dagger\} = E - \frac{3}{2}R - J_1 - J_2=0$, where $E$ denotes the scaling dimension of the corresponding operator. As usual, it is invariant under any continuous deformation of the theory.

For convenience, let us define new charges $Q^{I=1,2,\cdots,\,d}$ in terms of the $R$-charge and flavor charges as
\begin{equation}\label{eq:charge-map}
    \begin{aligned}
        &Q^I = \sum_{J=1}^{d-1}\xi^{IJ} f_J + \xi^{Id} R\ , \quad \xi^{Id} = \frac{1}{2} \qquad (I=1,2,\cdots,\,d)\ ,
    \end{aligned}
\end{equation}
with a $d\times d$ invertible matrix $\xi^{IJ}$ such that 
\begin{equation}
    [Q^I,\mathcal{Q}]=+\frac{1}{2}\mathcal{Q}\ ,
\end{equation}
and all the chiral multiplets in the theory carry rational $Q^I$'s. This is always possible for a gauge theory since the R-charge is constrained by superpotential and anomaly-free conditions, which are all linear.\footnote{We shall choose $\xi^{IJ}$ that makes the least common multiple of the denominators of each $Q^I$ for the chiral multiplets to be as small as possible. If possible, we will set all $Q^I$'s to be integers. This is possible for a large class of gauge theories including the ones obtained by D3-branes probing a Calabi-Yau cone over a Sasaki-Einstein 5-manifold.} Using the inverse matrix $\xi_{IJ}$, we find
\begin{equation}
    \begin{aligned}
        &f_I = \sum_{J=1}^d \xi_{IJ} Q^J\ , \quad 0=\sum_{J=1}^d\xi_{IJ} \qquad (I=1,2,\cdots,\,d-1)\ , \\
        &R = \sum_{J=1}^d \xi_{dJ} Q^J\ , \quad 2=\sum_{J=1}^d\xi_{dJ}\ ,
    \end{aligned}
\end{equation}
since $[f_I,\mathcal{Q}]=0$ and $[R,\mathcal{Q}]=+\mathcal{Q}$. Then, one can view $2Q^I$'s as various candidate $U(1)_R$ $R$-charges, where $\xi_{dJ} \;(J=1,2,\cdots, \, d)$ or the true superconformal $U(1)_R$-symmetry is determined by anomaly-free condition of the $R$-symmetry and the $a$-maximization procedure \cite{Intriligator:2003jj}. Now, the index \eqref{eq:tr-index1} can be rewritten as
\begin{equation}\label{eqn: tr-index}
    \mathcal{I}(\Delta,\sigma,\tau)=\Tr \left[ e^{2\pi i\sigma J_1}e^{2\pi i\tau J_2}\prod_{I=1}^de^{2\pi i\Delta_IQ^I} \right]\ ,
\end{equation}
with a constraint \cite{Choi:2018hmj, Kim:2019yrz, Amariti:2019mgp, Cassani:2021fyv}
\begin{equation}\label{eqn: constraints on chemical potentials (1)}
    \sum_{I=1}^d\Delta_I-\sigma-\tau=-1\ ,
\end{equation}
where $\Delta_I = \sum_{J=1}^{d-1} \delta^J \xi_{JI} + \delta \xi_{dI}$. From now on, we shall work with this form of the index. This particular form is technically easier to analyze and also accordant with the convention of 5d $\mathcal{N}=1$ gauged supergravity where $\Delta_I$'s correspond to various bulk gauge fields.

Let us now consider 4d $\mathcal{N}=1$ gauge theories, which flow to interacting SCFTs at IR.
Suppose the gauge theory has $n_\chi$ chiral multiplets $\Phi_\chi$'s in the representation $\mathcal{R}_\chi$ under the compact gauge group $G$ with the weights $\rho_\chi$. Then the superconformal index \eqref{eqn: tr-index} admits an integral representation given by \cite{Dolan:2008qi}
\begin{equation}\label{eqn: matrix integral-index}
    \mathcal{I}(\Delta,\sigma,\tau)=\frac{\kappa^{\textrm{rk}(G)}}{|W_G|}\oint_{\mathbb{T}^{\textrm{rk}(G)}}\prod_{i=1}^{\text{rk}(G)}du_i\;\frac{\prod_\chi\prod_{\rho_\chi\in\mathcal{R}_\chi}\Gamma\big(\rho_\chi(u)+\sum_{I=1}^dQ_\chi^I\Delta_I;\sigma,\tau\big)}{\prod_{\alpha\in \Delta_G}\Gamma\big(\alpha(u);\sigma,\tau\big)}\ ,
\end{equation}
where $\kappa\equiv (\sigma,\sigma)_\infty(\tau,\tau)_\infty$, and $Q^I_\chi$ denotes the $Q^I$-charge of the chiral multiplet $\Phi_\chi$. We will often call $N$ for the rank of the gauge group $G$ in this section. Here, the integration variable $u_i$'s parameterize the maximal torus of $G$, $\alpha$ parameterizes the roots of $G$, denoted by $\Delta_G$, and $|W_G|$ is the order of the Weyl group of $G$. The elliptic gamma function and the infinite $q$-Pochhammer symbol are defined as
\begin{equation}
\begin{aligned}
    \Gamma(z;\sigma,\tau)&=\prod_{m,n=0}^\infty\frac{1-e^{-2\pi iz}e^{2\pi i((m+1)\sigma+(n+1)\tau)}}{1-e^{2\pi iz}e^{2\pi i(m\sigma+n\tau)}}\ , \\
    (z;\tau)_\infty &= \prod_{n=0}^\infty (1-e^{2\pi i z} e^{2\pi i n \tau})\ .
    \end{aligned}
\end{equation}
Note that the index is well-defined only when $\Im\tau>0$ and $\Im\sigma>0$. 

Since $Q_\chi^I$'s are all rational numbers, \eqref{eqn: matrix integral-index} is invariant under the shifts of $\Delta_I$'s by the multiples of $p_I$'s, where $p_I$'s are the least common multiple of the denominators of each $Q^I$ for the chiral multiplets. In addition, it is invariant under arbitrary integer shifts of $\sigma$ and $\tau$. This is due to the periodicity of the elliptic gamma function
\begin{equation}
    \Gamma(z;\sigma,\tau)=\Gamma(z+1;\sigma,\tau)=\Gamma(z;\sigma+1,\tau)=\Gamma(z;\sigma,\tau+1)\ .
\end{equation}
We will distinguish the integer shifts of $\sigma,\tau$ and those of $\Delta_I$'s since two have different physical meanings in our large $N$ analysis. 
As we will see, our large $N$ saddle, which we shall introduce in Section \ref{sec:para}, only depends on $\sigma,\tau$ and is independent of $\Delta_I$'s. Since the explicit form of the large $N$ ansatz depends on $\sigma,\tau$, it spontaneously breaks the $\sigma,\tau$-shift symmetries of the matrix model. In other words, shifting $\sigma,\tau$ by arbitrary integers will generate new large $N$ solutions, as we will see in Section \ref{sec:more}.

On the other hand, given a particular large $N$ solution depending on $\sigma, \tau$, shifts of $\Delta_I$'s do not affect the matrix model. Thus, one can always perform any possible shifts of $\Delta_I$'s to make the large $N$ computations simple. This is analogous to the gauge fixing, which we now partially fix. We find that it is particularly convenient to analyze the large $N$ behavior of \eqref{eqn: matrix integral-index}
if shifted $\tilde\Delta_I \equiv \Delta_I + n_I p_I \; (n_I \in \mathbb{Z})$'s satisfy one of the following two choices: 
\begin{equation}\label{eq:constraint}
    \sum_{I=1}^d \tilde\Delta_I - \sigma - \tau = \mp 1\ .
\end{equation}
In other words, we choose either $\sum_{I=1}^d n_I p_I = 0,\,2$. While the case with the upper sign or $\sum_{I=1}^d n_I p_I = 0$ is always possible since we can set $n_I = 0$, the case with the lower sign or $\sum_{I=1}^d n_I p_I = 2$ may or may not be possible depending on $p_I$'s. In particular, if all $Q^I$'s are integer, \textit{i.e.} $p_I=1$, then both cases are possible. If there exists the residual shift symmetry of $\tilde\Delta_I$'s respecting the above conditions, it will be fixed later in Section \ref{sec:para}.
In fact, the above two cases, if both of them exist, are related by complex conjugation in the chemical potential space. If $(\tilde\Delta_I,\sigma,\tau)$ belongs to the upper case, $(-\tilde\Delta_I^*,-\sigma^*,-\tau^*)$ belongs to the lower case \cite{Choi:2021rxi}. Moreover, from \eqref{eqn: tr-index}, one can easily observe that
\begin{equation}
    \mathcal{I}(\tilde\Delta, \sigma, \tau)^* = \mathcal{I}(-\tilde\Delta^*, -\sigma^*, -\tau^*) \ .
\end{equation}
Therefore, in most cases, we will only analyze \eqref{eqn: matrix integral-index} under the upper case of \eqref{eq:constraint}, although the final results will incorporate both cases using the above complex conjugation. From now on, we will omit the $\tilde{\;}$ symbol from $\tilde\Delta_I$'s for simplicity.

To analyze the large $N (\equiv \textrm{rk} (G))$ behavior of \eqref{eqn: matrix integral-index}, we shall use the $SL(3,\mathbb{Z})$ modular property of the elliptic gamma function \cite{Felder_2000}:
\begin{equation}\label{eqn: SL(3,Z) identity (1)}
    \Gamma(z;\sigma,\tau)=e^{-\pi i P_+(z,\sigma,\tau)}\Gamma\left(-\frac{z+1}{\sigma};-\frac{1}{\sigma},-\frac{\tau}{\sigma}\right)\Gamma\left(\frac{z}{\tau};-\frac{1}{\tau},\frac{\sigma}{\tau}\right)\ ,
\end{equation}
\begin{equation}\label{eqn: SL(3,Z) identity (2)}
    \Gamma(z;\sigma,\tau)=e^{-\pi i P_-(z,\sigma,\tau)}\Gamma\left(-\frac{z}{\sigma};-\frac{1}{\sigma},-\frac{\tau}{\sigma}\right)\Gamma\left(\frac{z-1}{\tau};-\frac{1}{\tau},\frac{\sigma}{\tau}\right)\ ,
\end{equation}
where $P_\pm(z,\sigma,\tau)$ is given by
\begin{equation}\label{eqn: Cubic polynomials}
    \begin{split}
        P_\pm(z,\sigma,\tau)=&\frac{z^3}{3\sigma\tau}-\frac{\sigma+\tau\mp1}{2\sigma\tau}z^2+\frac{\sigma^2+\tau^2+3\sigma\tau\mp3\sigma\mp3\tau+1}{6\sigma\tau}z\\
        &\pm\frac{1}{12}(\sigma+\tau\mp1)\left(\frac{1}{\sigma}+\frac{1}{\tau}\mp1\right)\ .
    \end{split}
\end{equation}
The two identities are related to each other by complex conjugation by reparameterizing $(-z^*,-\sigma^*,-\tau^*)$ as $(z,\sigma,\tau)$. 

Using the $SL(3,\mathbb{Z})$ modular property \eqref{eqn: SL(3,Z) identity (1)}, the integrand of \eqref{eqn: matrix integral-index} can be reorganized as follows \cite{Choi:2021rxi}:
\begin{equation}\label{eqn: SL(3,Z) transformed matrix integral-index}
    \mathcal{I}(\Delta,\sigma,\tau) = \frac{\kappa^{\textrm{rk}(G)}}{|W_G|}\oint\prod_{i=1}^{\textrm{rk}(G)}du_i\;e^{-i\pi\mathbf{P}_+}e^{-V_\sigma(u)-V_\tau(u)}\ ,
\end{equation}
where the `potentials' $V_{\sigma,\tau} (u)$ are given by
\begin{align}\label{eqn: formal expression of matrix integral}
    \begin{split}
        -V_\sigma(u)&\equiv\sum_\chi\sum_{\rho_\chi\in\mathcal{R}_\chi}\log\Gamma\bigg(-\frac{\rho_\chi(u)+Q_\chi\cdot\Delta+1}{\sigma}; -\frac{1}{\sigma}, - \frac{\tau}{\sigma}\bigg) \\
        & \qquad \qquad -\sum_{\alpha\in\Delta_G}\log\Gamma\bigg(-\frac{\alpha(u)+1}{\sigma}; -\frac{1}{\sigma}, - \frac{\tau}{\sigma}\bigg)\ ,\\
        -V_\tau(u)&\equiv\sum_\chi\sum_{\rho_\chi\in\mathcal{R}_\chi}\log\Gamma\bigg(\frac{\rho_\chi(u)+Q_\chi\cdot\Delta}{\tau}; -\frac{1}{\tau},  \frac{\sigma}{\tau}\bigg) \\
        & \qquad \qquad -\sum_{\alpha\in\Delta_G}\log\Gamma\bigg(\frac{\alpha(u)}{\tau}; -\frac{1}{\tau},  \frac{\sigma}{\tau}\bigg)\ ,
    \end{split}
\end{align}
where $Q_\chi\cdot\Delta\equiv\sum_{I=1}^dQ_\chi^I\Delta_I$. 
The above functions are well-defined when the chemical potentials satisfy
\begin{equation}\label{eqn: restriction of modular parameters}
    \Im\left(-\frac{1}{\tau}\right)>0\ ,\qquad\Im\left(-\frac{1}{\sigma}\right)>0\ ,\qquad \Im\left(\frac{\sigma}{\tau}\right)>0\ ,
\end{equation}
While the first two conditions are automatically satisfied since the index is only well-defined when $\Im(\tau)>0$ and $\Im(\sigma)>0$, we will also require the third condition. One can perform the very same analysis with the roles of $\sigma$ and $\tau$ flipped, requiring $\Im\left(\frac{\sigma}{\tau}\right)<0$ instead. The collinear case $\Im\left(\frac{\sigma}{\tau}\right)=0$ can be studied through taking the collinear limit $\Im\left(\frac{\sigma}{\tau}\right) \to 0$ to the final results since the original integrand of \eqref{eqn: matrix integral-index} is smooth in that limit.\footnote{One may also study the large $N$ saddles of \eqref{eqn: matrix integral-index} precisely at $\textrm{Im}\left(\frac{\sigma}{\tau}\right)= 0$ as in \cite{Choi:2021rxi}.}

The `prefactor' $-i \pi\mathbf{P}_+$ of \eqref{eqn: SL(3,Z) transformed matrix integral-index} is given by
\begin{equation}
    -i\pi\mathbf{P}_\pm=-i\pi\left[\sum_\chi\sum_{\rho_\chi\in\mathcal{R}}P_\pm(\rho_\chi(u)+Q_\chi\cdot \Delta)-\sum_{\alpha\in\Delta_G}P_\pm(\alpha(u))\right]\ .
\end{equation}
where we included the both cases of \eqref{eq:constraint}.
Plugging in the definition \eqref{eqn: Cubic polynomials} of $P_\pm$, we can expand $-i\pi\mathbf{P}_\pm$ with respect to $u$ (the gauge holonomy variable) as follows:
\begin{itemize}
    \item $O(u^3)$:
    \begin{align}
        \hspace{-1.5cm}-\frac{i\pi}{3\sigma\tau}\left[\sum_\chi\sum_{\rho_\chi\in\mathcal{R}_\chi}\rho_\chi(u)^3+\sum_{\alpha\in\Delta_G}\alpha(u)^3\right]&=-\frac{i\pi}{3\sigma\tau}\left(d_3(G)+\sum_\chi d_3(\mathcal{R}_\chi)\right)|u|^3\nonumber\\ 
        &\equiv-\frac{i\pi}{3\sigma\tau}\Tr(GGG)|u|^3=0\ ,
    \end{align}

    \item $O(u^2)$:
    \begin{align}
        \begin{split}
            -&\frac{i\pi}{2\sigma\tau}\left[\sum_\chi\sum_{\rho_\chi\in\mathcal{R}_\chi}\big(2Q_\chi\cdot\Delta-(\sigma+\tau\mp1)\big)\rho_\chi(u)^2+\sum_{\alpha\in\Delta_G}\big(\sigma+\tau\mp1\big)\alpha(u)^2\right]\\
            &=-\frac{i\pi}{2\sigma\tau}\left[\sum_\chi\sum_{\rho_\chi\in\mathcal{R}_\chi}\left(\sum_{I=1}^d\big(2Q_\chi^I-1\big)\Delta_I\right)\rho_\chi(u)^2+\sum_{\alpha\in\Delta_G}\left(\sum_{I=1}^d\Delta_I\right)\alpha(u)^2\right]\\
            &=-\frac{i\pi}{2\sigma\tau}\sum_{I=1}^d\left[d_2(G)+\sum_\chi(2Q_\chi^I-1)d_2(\mathcal{R}_\chi)\right]\Delta_I|u|^2\\
            &\equiv-\frac{i\pi}{2\sigma\tau}\sum_{I=1}^{d}\Tr (Q^IGG )\Delta_I|u|^2 = 0\ ,
        \end{split}
    \end{align}

    \item $O(u^1)$:
    \begin{align}
        -&\hspace{-0.7cm}\frac{i\pi}{6\sigma\tau}\Bigg[\!\sum_\chi\!\sum_{\rho_\chi\in\mathcal{R}_\chi}\!\!\!\!\big(6(Q_\chi\cdot\Delta)^2-6(Q_\chi\cdot\Delta)(\sigma+\tau-1)+\sigma^2+\tau^2+3\sigma\tau\mp3\sigma\mp3\tau+1\big)\rho_\chi(u)\nonumber\\
        &\hspace{-0.7cm}\qquad\qquad+\sum_{\alpha\in\Delta_G}(\sigma^2+\tau^2+3\sigma\tau\mp3\sigma\mp3\tau+1)\alpha(u)\Bigg]\\
        &\hspace{-0.7cm}=-\frac{i\pi}{6\sigma\tau}\left[\sum_{I,J=1}^{d}\sum_\chi(2Q_\chi^I-1)(2Q_\chi^J-1)\sum_{\rho_\chi\in\mathcal{R}_\chi}\rho_\chi(u)+\sum_{\alpha\in\Delta_G}\alpha(u)\right]\Delta_I\Delta_J=0\nonumber \ .
    \end{align}

    \item $O(u^0)$:
    \begin{align}
        &\hspace{-0.8cm}\frac{-i\pi}{12\sigma\tau}\Bigg[\!\sum_\chi \!\!\sum_{\rho_\chi\in\mathcal{R}_\chi}\!\!\!\!\big(2Q_\chi\cdot\Delta-(\sigma+\tau\mp1)\big)\big(2(Q_\chi\cdot\Delta)^2-2(\sigma+\tau\mp1)(Q_\chi\cdot\Delta)+\sigma\tau\mp(\sigma+\tau)\big) \nonumber \\
        &\hspace{-0.8cm}\qquad\qquad+\sum_{\alpha\in\Delta_G}\big(\sigma+\tau\mp1\big)\big(\sigma\tau\mp(\sigma+\tau)\big)\Bigg]\Bigg|_{\sum_{I=1}^d\Delta_I-\sigma-\tau=\mp1} \nonumber \\
        &\hspace{-0.8cm}=-\frac{i\pi}{24\sigma\tau}\Bigg[\Bigg(\sum_{I,J,K=1}^{d}\sum_\chi\dim(\mathcal{R}_\chi)(2Q_\chi^I-1)(2Q_\chi^J-1)(2Q_\chi^K-1)+\dim(G)\Bigg)\Delta_I\Delta_J\Delta_K\nonumber\\
        &\hspace{-0.8cm}\qquad\qquad-(\sigma^2+\tau^2+1)\left(\sum_{I=1}^d\sum_\chi\dim(\mathcal{R}_\chi)(2Q_\chi^I-1)+\dim(G)\right)\Delta_I\Bigg] \\
        &\hspace{-0.8cm}\equiv-\frac{i\pi}{24\sigma\tau}\left[\sum_{I,J,K=1}^{d}\Tr( Q^IQ^JQ^K)\Delta_I\Delta_J\Delta_K-(\sigma^2+\tau^2+1)\sum_{I=1}^d\Tr(Q^I)\Delta_I\right]\nonumber\ ,
\end{align}
\end{itemize}
where we used \eqref{eq:constraint} and the following group theoretic identities:
\begin{align}\label{eqn: group identities}
    \begin{split}
        \sum_{\omega\in\mathcal{R}}\omega(u)^3&=d_3(\mathcal{R})\sum_{i,j,k=1}^{\textrm{rk}(G)}g_{ijk}u_iu_ju_k\equiv d_3(\mathcal{R})|u|^3\ ,\\
        \sum_{\omega\in\mathcal{R}}\omega(u)^2&=d_2(\mathcal{R})\sum_{i,j=1}^{\textrm{rk}(G)}h_{ij}u_iu_j\equiv d_2(\mathcal{R})|u|^2\ ,\\
        \sum_{\omega\in\mathcal{R}}\omega(u)&=0\ ,
    \end{split}
\end{align}
where the summation in the left hand side is over the weights $\omega$ of an irreducible representation $\mathcal{R}$ of the gauge group $G$. Here, $d_2(\mathcal{R})$ and $d_3(\mathcal{R})$ are the quadratic and cubic Dynkin index respectively, and the symbols $h_{ij}$ and $g_{ijk}$ are defined as
\begin{equation}
    h_{ij}=\tr\big(T_i T_j\big)\ ,\qquad g_{ijk}=\frac{1}{2}\tr\big(T_i\{T_j,T_k\}\big)\ ,
\end{equation}
where $T_i$'s are the Cartan generators of the Lie algebra of $G$ in the fundamental representation and $\{\cdot,\cdot\}$ denotes the anticommutator.
The $O(u^3)$ part of $-i\pi\mathbf{P}_\pm$ vanishes due to the absence of the gauge anomaly, and the $O(u^2)$ part vanishes due to the anomaly-free conditions for the $R$-symmetry and flavor symmetries, and finally the $O(u^1)$ vanishes due to the third identity of \eqref{eqn: group identities}.

Consequently, we have shown that
\begin{equation}\label{eq:prefactor}\hspace{-0.2cm}
    -i\pi \mathbf{P}_{\pm} = -\frac{i\pi}{24\sigma\tau}\left[\sum_{I,J,K=1}^{d}\Tr( Q^IQ^JQ^K)\Delta_I\Delta_J\Delta_K-(\sigma^2+\tau^2+1)\sum_{I=1}^d\Tr(Q^I)\Delta_I\right].
\end{equation}
Here, the trace anomaly coefficients completely govern the $u$-independent part of $-i\pi\mathbf{P}_\pm$. This fact was previously noticed in \cite{Gadde:2020bov, Jejjala:2022lrm}. In the Cardy-like limit, in which $\sigma, \tau \to 0$, the remaining integral of \eqref{eqn: SL(3,Z) transformed matrix integral-index} has a saddle point at $u_i=0$ with the vanishing saddle point value, thereby reproducing the 4d $\mathcal{N}=1$ Cardy formula \cite{Kim:2019yrz, Cabo-Bizet:2019osg, Cassani:2021fyv}. Note that so far we have not used any large $N$ approximation. The previous analysis simply shows that the prefactor $-i\pi\mathbf{P}_\pm$ will not affect our large $N$ saddle point problem. In other words, the large $N$ saddle points of the matrix integral \eqref{eqn: matrix integral-index} or \eqref{eqn: SL(3,Z) transformed matrix integral-index} are nothing but the extremums of the potentials \eqref{eqn: formal expression of matrix integral} in the large $N$ limit, which we now construct in the following subsection.

\subsection{Parallelogram ansatz}\label{sec:para}
In this subsection, we construct large $N$ saddle points of \eqref{eqn: SL(3,Z) transformed matrix integral-index}, \eqref{eqn: formal expression of matrix integral} employing the parallelogram ansatz \cite{Choi:2021rxi}. For concreteness, we consider a generic quiver gauge theory that has $n_v$ gauge nodes with $G_1,G_2,\cdots, G_{n_v}$ gauge groups, respectively, and hence the overall gauge group is given by $G=\prod_{a=1}^{n_v}G_a$. We shall analyze the large $N (\equiv \textrm{rk} (G))$ limit of the theory in which $\textrm{rk} (G_a) = O(N)$, and compute the free energy of the index \eqref{eqn: matrix integral-index} at the leading $O(N^2)$ order. The gauge nodes with $\textrm{rk} (G_a) = O(N^0)$ will give negligible contributions to the index at leading $O(N^2)$ order, which we dismiss. Each $G_a$, which is a compact simple Lie group, can be $SU(N_a)$, $SO(N_a)$, or $Sp(N_a)$.

Since we are interested in the large $N$ saddles, the gauge theories we consider should have a proper large $N$ limit. This restricts the matter contents in our theory to have rank-1 (fundamental) and rank-2 tensor representations (such as bifundamental, adjoint, (anti-)symmetric) under the gauge group. This is simply because that higher-rank representations result in IR-free (not asymptotically free) theory in large $N$. As it will become evident in this section, we also demand that the number of fundamental multiplets to be $O(N^0)$, not scaling with $N$. This excludes the Veneziano-like limit of taking $N_f/N$ to be fixed while taking $N$ large. From the AdS point of view, we fix the bulk gauge group while taking the large $N$ limit. We further demand that the $R$-charges for the elementary fields are all of order $O(N^0)$.\footnote{Usually the superconformal $R$-charges for the elementary fields scale as $O(N^0)$, which naturally gives a sparse spectrum of low-lying operators. However, there exists theories having elementary fields with $O(1/N)$ scaling of R-charges, which results in dense spectrum of low-lying operators in large $N$ \cite{Agarwal:2019crm, Agarwal:2020pol}. Our saddle is not applicable for those cases.} The full list of theories with a simple gauge group satisfying this condition has been classified in \cite{Agarwal:2020pol}. Consequently, these conditions suppress the contribution from the fundamentals (which is of $O(N^1)$) in our large $N$ matrix model (which is of $O(N^2)$) so that we can focus on the rank-2 tensor representations.

The chiral multiplet $\Phi_\chi$ in the rank-2 representation $\mathcal{R}_\chi$ can take the form of $(\mathbf{N}_a,\mathbf{N}_b)$, $(\overline{\mathbf{N}}_a,\overline{\mathbf{N}}_b)$, $(\mathbf{N}_a,\overline{\mathbf{N}}_b)$ or $(\overline{\mathbf{N}}_a,\mathbf{N}_b)$ under $G_a\times G_b$. Here $a=b$ case should be understood as the adjoint or (anti-)symmetric representation under $G_a$. Their weights $\rho_\chi(u)$ take one of the following forms:
\begin{equation}\label{eq:uab}
    +u_i^{(a)}+u_j^{(b)}\ , \;\; -u_i^{(a)}-u_j^{(b)}\ , \;\; +u_i^{(a)}-u_j^{(b)}\ , \;\; -u_i^{(a)}+u_j^{(b)}\ ,
\end{equation}
where we ignored the Cartan parts since they are independent of $u$.
We will collectively denote them as $u_{ij}^{(ab)}$. The sign choice will only be specified if needed. In fact, it is irrelevant in most of the analysis we carry out. Now, each chiral multiplet $\Phi_\chi$ contributes to the potentials \eqref{eqn: formal expression of matrix integral} as follows:
\begin{equation}\label{eqn: building blocks of potential}
\begin{aligned}
     & v_\sigma(u_{ij}^{(ab)}) \equiv -\sum_{i,j}\log\Gamma\left(-\frac{ u_{ij}^{(ab)}+Q_{\chi}\cdot \Delta+1}{\sigma};-\frac{1}{\sigma},-\frac{\tau}{\sigma}\right)\ ,\\
     & v_\tau(u_{ij}^{(ab)}) \equiv -\sum_{i,j}\log\Gamma\left(\frac{ u_{ij}^{(ab)}+Q_{\chi}\cdot\Delta}{\tau};-\frac{1}{\tau},\frac{\sigma}{\tau}\right)\ ,
     \end{aligned}
\end{equation}
where the summation range of $i,j$ should be appropriately chosen according to the representation $\mathcal{R}_\chi$. Contribution from the vector multiplet of $a$'th gauge node can also be expressed using the above formulae with overall minus sign, and inserting $u_{ij}^{(aa)} = \alpha_a(u)$ and $Q_\chi=(0,0,\cdots,0)$, where the roots $\alpha_a$ of $G_a$ also takes one of the form of \eqref{eq:uab}.

\begin{figure}\hspace{-1.5cm}
    \centering
    \begin{subfigure}[b]{0.48\linewidth}
        \hspace{2cm}\includegraphics[width=0.7\linewidth]{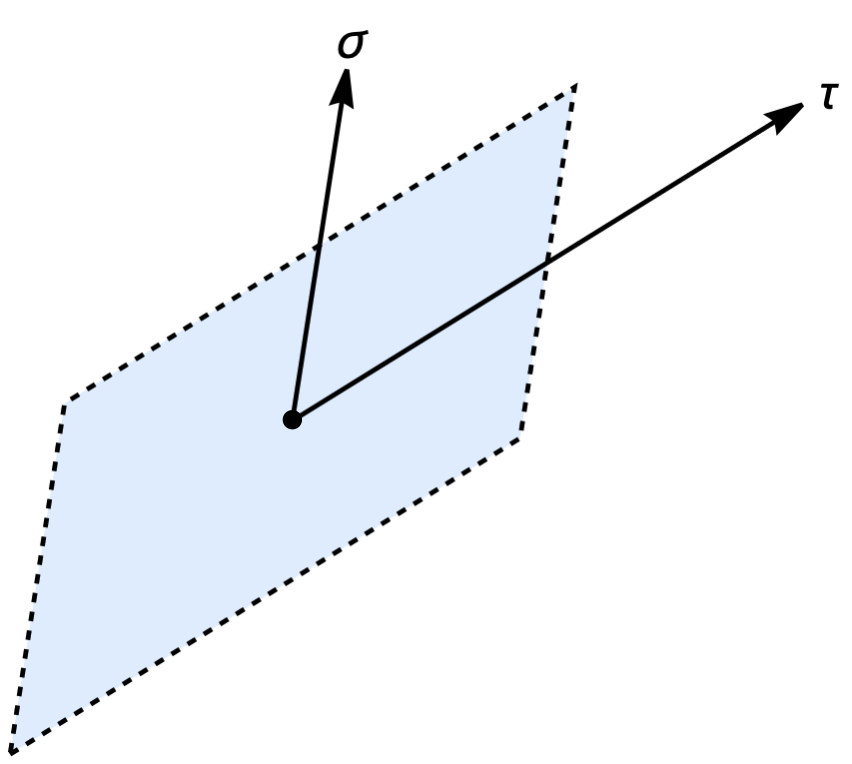}
        \caption{$SU(N)$}
        \label{fig:A}
    \end{subfigure}
    \hspace{-1cm}
    \begin{subfigure}[b]{0.48\linewidth}    
        \hspace{2cm}
        \includegraphics[width=0.7\linewidth]{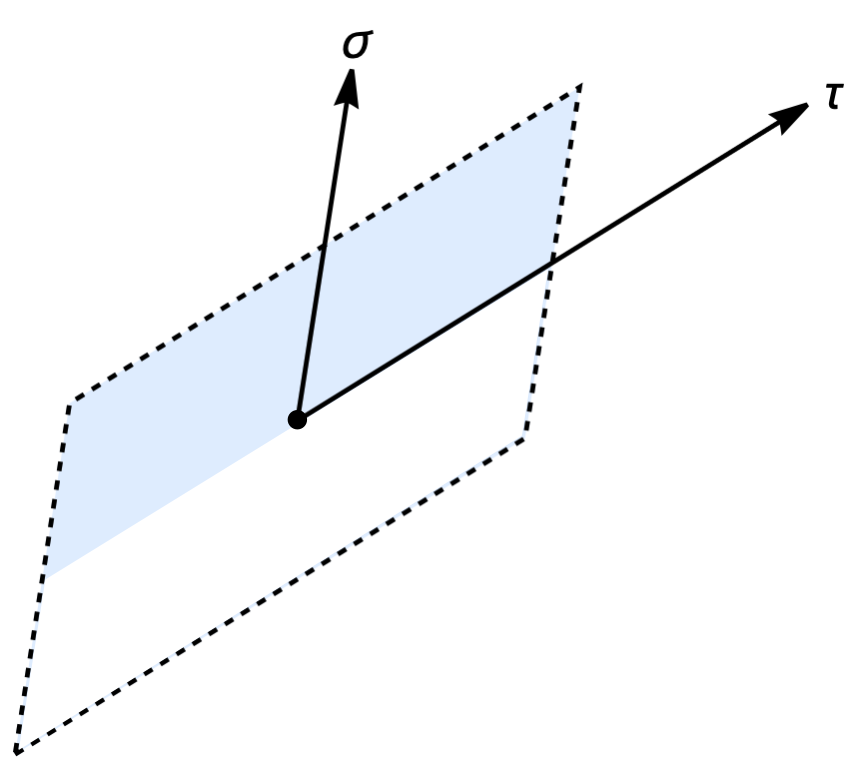}
        \caption{$SO(2N)$, $SO(2N+1)$, $Sp(N)$}
        \label{fig:B}
    \end{subfigure}
    \caption{$N$ eigenvalues are uniformly distributed in the colored region of the complex $u$ plane.}
    \label{fig: Parallelogram ansatz}
\end{figure}

We now introduce our uniform parallelogram ansatz in the large $N$ limit. Specific form of the ansatz for each gauge node depends on the type of the gauge group. For $SU(N_a)$ gauge nodes, the ansatz is given as \cite{Choi:2021rxi}
\begin{equation}\label{eqn: parallelogram ansatz (2)}
    u_i^{(a)}=\sigma x_i+\tau y_i\ ,\qquad x_i,y_i\in\left(-\frac{1}{2},\frac{1}{2}\right)\ ,
\end{equation}
with $\sum_{i=1}^{N_a} u_i^{(a)} = 0$, and for $SO(2N_a)$, $SO(2N_a+1)$ or $Sp(N_a)$ gauge nodes, it is given by
\begin{equation}\label{eqn: parallelogram ansatz (3)}
    u_i^{(a)}=\sigma x_i+\tau y_i\ ,\qquad x_i\in\left(0,\frac{1}{2}\right)\, , \;\; y_i\in\left(-\frac{1}{2},\frac{1}{2}\right)\ ,
\end{equation}
where the eigenvalue $u_{i=1,2,\cdots, N_a}^{(a)}$'s are evenly distributed in the parallelogram for both cases as depicted in Figure \ref{fig: Parallelogram ansatz}. Note that we assumed $\textrm{Im} \left(\frac{\sigma}{\tau}\right)>0$ in \eqref{eqn: restriction of modular parameters}. Next, let us introduce a $2$-dimensional eigenvalue density $\rho^{(a)}(x,y)$ given by
\begin{equation}
    \rho^{(a)}(x,y)=\frac{1}{N_a}\sum_{i=1}^{N_a}\delta(x-x_i)\delta(y-y_i)\ , \qquad \int dxdy\rho^{(a)}(x,y)=1\ ,
\end{equation}
where the integration range is either \eqref{eqn: parallelogram ansatz (2)} or \eqref{eqn: parallelogram ansatz (3)}.
In the large $N_a$ limit, eigenvalues are densely distributed in the parallelogram \eqref{eqn: parallelogram ansatz (2)}, \eqref{eqn: parallelogram ansatz (3)}, so we can regard $\rho^{(a)} (x,y)$ as a continuous function. In particular, since eigenvalues are evenly distributed, 
\begin{equation}
\rho^{(a)}(x,y) \to \left\{ \begin{array}{ll}
1\ ,\qquad \qquad & SU(N_a) \\
2\ ,\qquad \qquad & SO(2N_a),\; SO(2N_a+1), \;Sp(N_a)
\end{array} \right.\ ,
\end{equation}
inside the parallelogram \eqref{eqn: parallelogram ansatz (2)}, \eqref{eqn: parallelogram ansatz (3)}, in the large $N_a$ continuum limit. Then, the summation over the eigenvalues can be approximated by the integration over $x,y$ with the uniform areal eigenvalue density $\rho^{(a)}(x,y)$ as follows:
\begin{align}
    \sum_{i=1}^{N_a} f(u^{(a)}_i) \;&\rightarrow\;N_a\int_{-1/2}^{1/2}\int_{-1/2}^{1/2}dx dy \; f(\sigma x + \tau y) +O(N_a^0) \ ,
\end{align}
for $SU(N_a)$ gauge nodes, and
\begin{align}
    \sum_{i=1}^{N_a} \left(f(u^{(a)}_i) + f(-u^{(a)}_i) \right) \;&\rightarrow\;2N_a\int_{-1/2}^{1/2}\int_{-1/2}^{1/2}dx dy \; f(\sigma x + \tau y) +O(N_a^0) \ ,
\end{align}
for $SO(2N_a), \, SO(2N_a+1), \, Sp(N_a)$ gauge nodes. Here, we used the fact that there always exist pairs of positive/negative weights for $SO(2N_a), \, SO(2N_a+1), \, Sp(N_a)$ gauge groups.

To show that the parallelogram ansatz \eqref{eqn: parallelogram ansatz (2)}, \eqref{eqn: parallelogram ansatz (3)} indeed represents large $N$ saddle points of \eqref{eqn: SL(3,Z) transformed matrix integral-index}, we shall verify that each $v_\sigma(u_{ij}^{(ab)})$ and $v_\tau(u_{ij}^{(ab)})$ give vanishing force at the leading order $O(N^1)$ within the parallelogram region \eqref{eqn: parallelogram ansatz (2)}, \eqref{eqn: parallelogram ansatz (3)}.\footnote{In principle, it is possible that although individual forces do not vanish, they cancel each other so that the total force does. Here, we assume that we can find an appropriate shift of $\Delta_I$'s, using the residual shift symmetry respecting \eqref{eq:constraint}, which makes each force vanish.} 
Taking a derivative of the potential with respect to a particular eigenvalue $u_i^{(a)}$, we obtain
\begin{equation}
    \sum_{j}\bigg[\frac{\partial}{\partial u_i^{(a)}}v_\sigma\big( u_{ij}^{(ab)}\big)+\frac{\partial}{\partial u_i^{(a)}}v_\tau\big( u_{ij}^{(ab)}\big)\bigg]= \pm \sum_{j}\bigg[\frac{1}{\sigma}\frac{\partial}{\partial x_j}v_\sigma\big(u_{ij}^{(ab)}\big)+\frac{1}{\tau}\frac{\partial}{\partial y_j}v_\tau\big( u_{ij}^{(ab)}\big)\bigg]\ ,
\end{equation}
where we omitted possible overall normalization factor, which is irrelevant in the following discussion. Here, we used $\frac{\partial}{\partial u_i^{(a)}} = \pm\frac{\partial}{\partial u_j^{(b)}}$ and $u_j^{(b)}=\sigma x_j+\tau y_j$ so that the derivatives can be replaced by either $\frac{1}{\sigma}\frac{\partial}{\partial x_j}$ or $\frac{1}{\tau}\frac{\partial}{\partial y_j}$. The $\pm$ sign is fixed once we use one of the explicit expressions in \eqref{eq:uab} for $u_{ij}^{(ab)}$. In the large $N$ continuum limit, the above force becomes
\begin{equation}
\begin{aligned}   
    &N_b\int_{-1/2}^{1/2}\int_{-1/2}^{1/2}dx_2dy_2\left(\frac{1}{\sigma}\frac{\partial}{\partial x_2}v_\sigma\big( \sigma x_{12} + \tau y_{12}\big)+\frac{1}{\tau}\frac{\partial}{\partial y_2}v_\tau\big( \sigma x_{12} + \tau y_{12}\big)\right)\ , \\
    =\;&\frac{N_b}{\sigma}\int_{-1/2}^{1/2}dy_2 \; v_\sigma\big(\sigma x_{12} + \tau y_{12}\big)\Big|^{x_2 = +1/2}_{x_2 = -1/2} + \frac{N_b}{\tau}\int_{-1/2}^{1/2}dx_2v_\tau\big(\sigma x_{12} + \tau y_{12}\big)\Big|^{y_2 = +1/2}_{y_2 = -1/2}\ ,
\end{aligned}
\end{equation}
where $x_{12}$ can be one of $x_1+x_2$, $-x_1-x_2$, $x_1-x_2$, $-x_1+x_2$, and so does $y_{12}$, in accordance with $u_{ij}^{(ab)}$.
We see that both terms vanish when $v_\sigma$ and $v_\tau$ are periodic in $\sigma$ and $\tau$ respectively. Namely, the force-free condition is satisfied if
\begin{equation}\label{eqn: force-free condition}
        v_\sigma( z+ \sigma) = v_\sigma(z)\ , \quad v_\tau( z+ \tau) = v_\tau(z) \ .
\end{equation}
From \eqref{eqn: building blocks of potential}, it is true that $e^{v_\sigma}$ and $e^{v_\tau}$ are indeed periodic in $\sigma, \tau$ respectively, due to the periodicity of $\Gamma$. When we have a fundamental chiral multiplet, the force-free condition is not met by our saddle. However, these terms are always subleading in $N$ (unless we take Veneziano-like limit of scaling the number of fundamentals) so that we ignore such contribution in the large $N$ limit.

Now it seems like we are done showing that our ansatz is indeed a saddle point of the integral. 
However, the periodicity of $v_\sigma$ and $v_\tau$ can be ruined if there exist a logarithmic branch point inside the integral domain \eqref{eqn: parallelogram ansatz (2)}, \eqref{eqn: parallelogram ansatz (3)}. Fortunately, we find that such branch points are absent in certain region of the parameter space for $\Delta_I, \sigma, \tau$, which we shall study now.
Let us first analyze the contribution from the chiral multiplets. Absence of the branch points of $v_\sigma$ in the integral domain requires that the following elliptic gamma function
\begin{align}
    \Gamma\left(-\frac{u_{12}+Q_{\chi}\cdot\Delta+1}{\sigma};-\frac{1}{\sigma},-\frac{\tau}{\sigma}\right)=\prod_{m,n=0}^\infty\frac{1-e^{2\pi i\left(\frac{u_{12}+Q_{\chi}\cdot\Delta+1}{\sigma}-\frac{m+1}{\sigma}-\frac{(n+1)\tau}{\sigma}\right)}}{1-e^{2\pi i\left(-\frac{u_{12}+Q_{\chi}\cdot\Delta+1}{\sigma}-\frac{m}{\sigma}-\frac{n\tau}{\sigma}\right)}}
\end{align}
has no zeros/poles from numerators/denominators for $u_{12} = \sigma x_{12} + \tau y_{12}$ in the region \eqref{eqn: parallelogram ansatz (2)} or \eqref{eqn: parallelogram ansatz (3)}. To avoid poles from the denominator, we require
\begin{equation}
    \Bigg|\exp\left[2\pi i\left(-\frac{u_{12}+Q_{\chi}\cdot\Delta+1}{\sigma}-\frac{m}{\sigma}-\frac{n\tau}{\sigma}\right)\right]\Bigg|<1\ ,\qquad \forall m,n \geq 0\ .
\end{equation}
Note that $\Im\left[-\frac{u_{12}+Q_{\chi}\cdot\Delta+1}{\sigma}-\frac{m}{\sigma}-\frac{n\tau}{\sigma}\right]\geq \Im\left(-\frac{Q_{\chi}\cdot\Delta+1-\tau}{\sigma}\right)$, where the inequality is saturated at $m=n=0$ and $y_{12}=-1$. Hence, the above inequality holds if and only if 
\begin{equation}
\Im\left(\frac{Q_{\chi}\cdot\Delta+1-\tau}{\sigma}\right)<0\ .
\end{equation}
One can make a similar analysis for the other terms in $v_\sigma$ and $v_\tau$, which yields similar constraints as follows:
\begin{align}\label{eqn: four-chemical potential inequalities}
    \begin{split}
        &\Im\left(\frac{Q_{\chi}\cdot\Delta}{\sigma}\right)>0, \quad\Im\left(\frac{1-\tau+Q_{\chi}\cdot\Delta}{\sigma}\right)<0\ , \\
        &\Im\left(\frac{1+Q_{\chi}\cdot\Delta}{\tau}\right)<0\ , \quad \Im\left(\frac{-\sigma+Q_{\chi}\cdot\Delta}{\tau}\right)>0\ .
    \end{split}
\end{align}

Next, let us analyze the contribution of the vector multiplets. Firstly, the integrand of $v_\sigma (\alpha_a(u))$ takes the following form:
\begin{equation}
     \sum_{m,n\geq 0} \left[-\log \left(1-e^{2\pi i\left(-\frac{\alpha_a(u)+1}{\sigma}-\frac{m}{\sigma}-\frac{n\tau}{\sigma}\right)}\right) + \log \left(1-e^{2\pi i\left(\frac{\alpha_a(u)+1}{\sigma}-\frac{m+1}{\sigma}-\frac{(n+1)\tau}{\sigma}\right)}\right)\right]\ ,
\end{equation}
where the first term is periodic when $\Im\left(\frac{1-\tau}{\sigma}\right)<0$, which is true once we assumed \eqref{eqn: four-chemical potential inequalities}, and hence it does not induce any force. The force from the second term is explicitly given by
\begin{equation}\label{vec-force}
    \sum_{m,n\geq 0} \int^{1/2}_{-1/2} \int^{1/2}_{-1/2} \frac{dx_2 dy_2}{1-e^{-2\pi i \left[x_{12}+\frac{1}{\sigma}(\tau (y_{12}-n-1) -m )  \right] }}\ ,
\end{equation}
where we dropped the overall constant. Except for the case with $m=n=0$, the following always holds:
\begin{equation}
\Im \left[x_{12}+\frac{1}{\sigma}(\tau (y_{12}-n-1) -m ) \right] >0\ .
\end{equation}
Thus, the nontrivial contribution to \eqref{vec-force} can only come from the case when $m=n=0$. Even in that case, if $y_{12} = y_1  \pm y_2$, the above inequality holds unless $y_1 = \frac{1}{2}$. When $y_{12} = - y_1 \pm y_{2}$, it holds unless $y_1 = - \frac{1}{2}$. Therefore, the only (seemingly) non-vanishing force takes the following form:
\begin{equation}
    \int^{1/2}_{-1/2} \int^{1/2}_{-1/2} \frac{dx_2 dy_2}{1-e^{-2\pi i \left[x_{2}+\frac{\tau}{\sigma} (y_{2}-\frac{1}{2})  \right] }}\ ,
\end{equation}
where we appropriately redefined $x_2$, $y_2$ and dropped the overall constant. The integrand has a pole at $x_2=0, \, y_2=\frac{1}{2}$. So we divide the integral domain into two regions: $\epsilon$-neighborhood of the pole and the remaining region. The integral in the latter vanishes since the integrand is analytic in that region. The absolute value of the integral in the $\epsilon$-neighborhood of the pole is given by
\begin{equation}
\begin{aligned}
    \left|\int^{\epsilon}_{-\epsilon} dx_2 \int^{0}_{-\epsilon} \frac{dy_2}{1-e^{-2\pi i \left[x_{2}+\frac{\tau}{\sigma} y_{2}  \right]}} \right| & \leq \int^{\epsilon}_{-\epsilon} dx_2 \int^{0}_{-\epsilon} \frac{dy_2}{\left|2\pi i  \left(x_2 + \frac{\tau}{\sigma} y_2 \right)  \right|} \left(1+ O(\epsilon)\right) \\
    &=O(\epsilon \log \epsilon)\ ,
    \end{aligned}
\end{equation}
which vanishes in the limit $\epsilon \to 0$. Therefore, the force from $v_\sigma (\alpha_a(u))$ part of the vector multiplets totally vanishes.

Finally, let us analyze the force from $v_\tau (\alpha_a(u))$ part of the vector multiplets whose integrand takes the following form:
\begin{equation}\label{vec-v}
    \sum_{m,n \geq 0} \left[\log \left(1-e^{2\pi i\left(-\frac{\alpha_a(u)}{\tau}-\frac{m+1}{\tau}+\frac{(n+1)\sigma}{\tau}\right)}\right) - \log \left(1-e^{2\pi i\left(\frac{\alpha_a(u)}{\tau}-\frac{m}{\tau}+\frac{n\sigma}{\tau}\right)}\right)\right]\ .
\end{equation}
The first term is again periodic since $\Im \left( \frac{1}{\tau} \right)<0$, hence does not induce any force. Except for the case when $m=n=0$, the second term is also periodic. However, when $m=n=0$, there indeed exist branch points at $\alpha_a(u)=0$. From now on, we shall separate the problematic Haar-measure-like factor, $m=n=0$ term of the second term of \eqref{vec-v}, from each gauge node, and redefine $V_\tau(u)$ as
\begin{equation}
    e^{-V_\tau(u)} \quad \to \quad \prod_{a=1}^{n_v}\prod_{\alpha_a\in \Delta_{G_a}}\left(1-e^{2\pi i\frac{\alpha_a(u)}{\tau}}\right)\cdot e^{-\Tilde{V}_\tau(u)}\ .
\end{equation}
Consequentially, the matrix integral of the index \eqref{eqn: SL(3,Z) transformed matrix integral-index} becomes
\begin{equation}\label{eq:final_integral}
    \mathcal{I}(\Delta,\sigma,\tau) = \frac{\kappa^{\textrm{rk}(G)}}{|W_G|}\oint\prod_{i=1}^{\textrm{rk}(G)}du_i\prod_{a=1}^{n_v}\prod_{\alpha_a\in \Delta_{G_a}}\left(1-e^{2\pi i\frac{\alpha_a(u)}{\tau}}\right) e^{-i\pi\mathbf{P}_+}e^{-V_\sigma(u)-\Tilde{V}_\tau(u)}\ ,
\end{equation}
where $V_\sigma$ and $\Tilde{V}_\tau$ satisfy all the required periodicity \eqref{eqn: force-free condition} for vanishing force in the parallelogram \eqref{eqn: parallelogram ansatz (2)}, \eqref{eqn: parallelogram ansatz (3)} under the assumption that the chemical potentials satisfy \eqref{eqn: four-chemical potential inequalities}.

We next introduce the following integral identity 
\begin{align}\label{eqn: integral identity}
    \begin{split}
        \frac{1}{|W_G|}&\oint_{\mathbb{T}^{\textrm{rk}(G)}}\prod_{i=1}^{\textrm{rk}(G)}du_i\prod_{\alpha \in \Delta_G^+}(1-e^{2\pi i\gamma \alpha(u)})(1-e^{-2\pi i\gamma \alpha(u)})\cdot f(u) \\ 
        =&\oint_{\mathbb{T}^{\textrm{rk}(G)}}\prod_{i=1}^{\textrm{rk}(G)}du_i\prod_{\alpha \in \Delta_G^+}(1-e^{2\pi i\gamma \alpha(u)})\cdot f(u)\ ,
    \end{split}
\end{align}
where $\Delta_G^+$ denotes the set of all positive roots in $G$.
This identity holds for any constant $\gamma$ and Weyl-invariant function $f(u)$ for a compact semisimple Lie group $G$. When $\gamma=1$, the right-hand side of this identity is widely used to reduce the computation complexity involving the Haar measure to the one only with half of the Haar measure. 
Our main claim is that the parallelogram ansatz \eqref{eqn: parallelogram ansatz (2)}, \eqref{eqn: parallelogram ansatz (3)} solves the large $N$ saddle point problem after using the above identity to \eqref{eq:final_integral}. To justify this, let us apply the identity with $\gamma=\frac{1}{\tau}$ as follows:
\begin{equation}\label{eq:mod-integral}
    \mathcal{I}(\Delta,\sigma,\tau) = \kappa^{\textrm{rk}(G)}\oint\prod_{i=1}^{\textrm{rk}(G)}du_i\prod_{a=1}^{n_v}\prod_{\alpha_a\in \Delta^+_{G_a}}\left(1-e^{2\pi i\frac{\alpha_a(u)}{\tau}}\right)e^{-i\pi\mathbf{P}_+}e^{-V_\sigma(u)-\Tilde{V}_\tau(u)}\ .
\end{equation}
As an additional part of ansatz \eqref{eqn: parallelogram ansatz (2)}, \eqref{eqn: parallelogram ansatz (3)}, we order the eigenvalue $u_i^{(a)}$'s such that \cite{Choi:2021rxi}
\begin{equation}\label{eq:order}
i<j \quad \Rightarrow \quad x_i > x_j\ .
\end{equation}
Then, the half-Haar-measure-like factor has no zeros since
\begin{equation}
        \left|e^{2\pi i\frac{\alpha_a(u)}{\tau}}\right| \ni \left|e^{2\pi i\frac{\sigma (x_i-x_j) +\tau (y_i-y_j)}{\tau}}\right| = \left|e^{2\pi i\frac{\sigma}{\tau}(x_i-x_j)}\right| < 1\
\end{equation}
for positive roots ($i<j$) of $SU(N_a)$ in \eqref{eqn: parallelogram ansatz (2)}, and
\begin{equation}
        \left|e^{2\pi i\frac{\alpha_a(u)}{\tau}}\right| \ni \left\{ \begin{array}{ll}
\left|e^{2\pi i\frac{\sigma (x_i-x_j) +\tau (y_i-y_j)}{\tau}}\right| = \left|e^{2\pi i\frac{\sigma}{\tau}(x_i-x_j)}\right| < 1 \\
\left|e^{2\pi i\frac{\sigma (x_i+x_j) +\tau (y_i+y_j)}{\tau}}\right| = \left|e^{2\pi i\frac{\sigma}{\tau}(x_i+x_j)}\right| < 1
\end{array} \right.
\end{equation}
for positive roots ($i<j$) of $SO(2N_a)$, $SO(2N_a +1)$, $Sp(N_a)$ in \eqref{eqn: parallelogram ansatz (3)}, when $\textrm{Im} \left(\frac{\sigma}{\tau}\right)>0$. Namely, the half-Haar-measure-like factor does not yield branch points to the potential of the index matrix model \eqref{eq:mod-integral}. This completes the proof that our `ordered' parallelogram ansatz \eqref{eqn: parallelogram ansatz (2)}, \eqref{eqn: parallelogram ansatz (3)} indeed solves the large $N$ saddle point equation of the index matrix integral \eqref{eq:mod-integral} under the assumptions \eqref{eq:constraint}, \eqref{eqn: restriction of modular parameters}, \eqref{eqn: four-chemical potential inequalities}.

Another consequence of the above analysis is that the potentials (including the contribution from the half-Haar-measure-like factor) of \eqref{eq:mod-integral} are 
periodic in $\sigma$ or $\tau$ and holomorphic in the ordered parallelogram region \eqref{eqn: parallelogram ansatz (2)}, \eqref{eqn: parallelogram ansatz (3)}. Therefore, upon evaluating the periodic integral in $\sigma$ or $\tau$ direction, they vanish at the parallelogram saddle point in the large $N$ continuum limit by Cauchy's integral theorem in the leading $O(N^2)$ order. Only the prefactor $\mathbf{P}_+$ contributes to the large $N$ free energy of the index \eqref{eq:mod-integral}.

Let us mention the case when our ansatz fails to become a valid saddle. When we have elementary fields whose $R$-charges scale as $O(1/N)$, the first condition of \eqref{eqn: four-chemical potential inequalities} can never be satisfied. This is because, in the large $N$ limit, we encounter dangerous terms of the form $\left(1-e^{2\pi i \frac{u_{ij}}{\tau}}\right)^{-1}$ just like the case of vector multiplet. However, in this case, such a dangerous term appears in the denominator, which cannot be removed using the identity \eqref{eqn: integral identity} involving the Haar-measure-like factor. Thus, our ansatz does not solve the saddle point equation for the theories with `dense spectrum' \cite{Agarwal:2019crm, Agarwal:2020pol}, which includes (the $\CN=1$ gauge theory description of) a certain large $N$ limit of Argyres-Douglas theories \cite{Maruyoshi:2016aim, Agarwal:2016pjo}.

In summary, we have demonstrated that the uniform parallelogram ansatz \eqref{eqn: parallelogram ansatz (2)} and \eqref{eqn: parallelogram ansatz (3)} with the ordering \eqref{eq:order} exactly solves the matrix model for the index \eqref{eq:mod-integral} (and thus \eqref{eqn: matrix integral-index}) in the large $N$ limit, provided that the chemical potentials satisfy
\begin{align}\label{eq:chem-con1}
    \begin{split}
        &\sum_{I=1}^d \Delta_I - \sigma - \tau = - 1\ , \qquad \textrm{Im} \left(\frac{\sigma}{\tau}\right)>0\ , \\
        &\textrm{Im}\left(\frac{1+\sigma}{\tau}\right)<\Im\left(\frac{1+Q_{\chi}\cdot\Delta}{\tau}\right)<0<\Im\left(\frac{Q_{\chi}\cdot\Delta}{\sigma}\right)<\textrm{Im}\left(\frac{-1+\tau}{\sigma}\right)\ .
    \end{split}
\end{align}
As explained previously, the case with $\textrm{Im} \left(\frac{\sigma}{\tau}\right)<0$ can be studied by flipping the role of $\sigma$ and $\tau$ of the above case. If there exists the conjugate sector, it can be studied similarly where the chemical potentials satisfy
\begin{align}\label{eq:chem-con2}
    \begin{split}
        &\sum_{I=1}^d \Delta_I - \sigma - \tau = + 1\ , \qquad \textrm{Im} \left(\frac{\sigma}{\tau}\right)>0\ , \\
        &\textrm{Im}\left(\frac{1+\sigma}{\tau}\right)<\Im\left(\frac{Q_{\chi}\cdot\Delta}{\tau}\right)<0<\Im\left(\frac{-1+Q_{\chi}\cdot\Delta}{\sigma}\right)<\textrm{Im}\left(\frac{-1+\tau}{\sigma}\right)\ ,
    \end{split}
\end{align}
or its $\sigma \leftrightarrow \tau$ flipped version. One can also take the collinear limit, in which $\textrm{Im} \left(\frac{\sigma}{\tau}\right)\to 0$, from both cases smoothly. If both sectors \eqref{eq:chem-con1}, \eqref{eq:chem-con2} exist, upon Legendre transformation of the free energy to the microcanonical ensemble, they give complex conjugate contributions, thereby realizing the oscillating signs of the index degeneracy \cite{Agarwal:2020zwm, Choi:2020baw}. The constraints \eqref{eq:chem-con1}, \eqref{eq:chem-con2} are reminiscent of the stability conditions of the Euclidean BPS black hole solutions against the condensation of the D3-brane instantons \cite{Aharony:2021zkr}. It would be interesting to study the relation between them if any.

In any case, the resulting large $N$ free energy of the index \eqref{eqn: matrix integral-index} is given by\footnote{This formula looks very similar to the supersymmetric Casimir energy studied in \cite{Bobev:2015kza,Cabo-Bizet:2018ehj}. It would be desirable to clarify the relation between them.}
\begin{align}\label{eq:free}
    \log\mathcal{I} &= -\frac{i\pi}{24\sigma\tau}\left[\sum_{I,J,K=1}^{d}\!\!\Tr( Q^IQ^JQ^K)\Delta_I\Delta_J\Delta_K-(\sigma^2+\tau^2+1)\sum_{I=1}^d\Tr(Q^I)\Delta_I\right] + O(N) \nonumber\\
    & \sim -\frac{i\pi}{24\sigma\tau}\sum_{I,J,K=1}^{d}\Tr( Q^IQ^JQ^K)\Delta_I\Delta_J\Delta_K\ ,
    \end{align}
where we only kept terms of order $O(N^2)$ from the prefactor $-i\pi \mathbf{P}_\pm$. 
We can drop the second term in the first line of \eqref{eq:free} for the gauge theories of our interest, where $\textrm{rk} (G_a) = O(N)$, since one can show that $\Tr(Q^I) = O(N^1)$ or $O(N^0)$ at large $N$ as follows:
\begin{equation}
    \begin{aligned}
        \Tr (Q^I)& = \frac{1}{2}\sum_a \left[ \dim (G_a) + \sum_{\chi_a} (2Q^I_{\chi_a} -1) \dim (\mathcal{R}_{\chi_a})\right] \\
        & =\frac{1}{2} \sum_a \textrm{rk}(G_a) \left(  d_2(G_a)+\sum_{\chi_a}(2Q_{\chi_a}^I-1)d_2(\mathcal{R}_{\chi_a}) \right) + O(N    ) \\
        &=  \sum_a \textrm{rk}(G_a) \Tr  (Q^I G_aG_a) + O(N) \\
        &= O(N)\ , 
    \end{aligned}
\end{equation}
where the second line comes from the fact that for the rank-2 tensor representations $d_2(\mathcal{R}) / \dim(\mathcal{R}) \sim 1/N$ at large $N$. In the third line, we find the leading order term vanishes since $Q^I$-symmetry is anomaly-free.\footnote{Similar argument was used to show that certain $\CN=1$ gauging of Argyres-Douglas theories have vanishing $\Tr R$ so that they have $a=c$ \cite{Kang:2021ccs}. We thank Ki-Hong Lee for the proof.}

As a concrete holographic example, let us consider the 4d $\CN=1$ SCFT describing the low energy dynamics of a stack of $N$ D3-branes probing a conical Calabi-Yau 3-fold singularity. This theory is dual to type IIB string theory on AdS$_5 \times$ SE$_5$, where SE$_5$ is the Sasaki-Einstein 5-manifold serving as the base of the Calabi-Yau 3-fold. Then, the above formula can be written as
\begin{equation}\label{eqn: free energy}
    \begin{split}
        \log\mathcal{I}(\Delta,\sigma,\tau)\sim -\frac{i\pi N^2}{24\sigma\tau}\sum_{I,J,K=1}^{d}C^{IJK}\Delta_I\Delta_J\Delta_K\ ,
    \end{split}
\end{equation}
where $C^{IJK} \equiv \frac{1}{N^2} \Tr(Q^IQ^JQ^K)$'s correspond to the Chern-Simons couplings of the 5d $\CN=1$ gauged supergravity, which can be obtained from a consistent Kaluza-Klein truncation of type IIB supergravity on $\textrm{SE}_5$. This precisely matches the entropy function of the BPS black holes in AdS$_5 \times \textrm{SE}_5$ \cite{Hosseini:2017mds, Hosseini:2018dob,Cabo-Bizet:2018ehj,Cassani:2019mms,Lanir:2019abx,Amariti:2019mgp,Benini:2020gjh}, thus accounting for their microstates.\footnote{In fact, the terms of order $O(N^1)$ including $\textrm{Tr}(Q)$ in the first line of \eqref{eq:free} match the four-derivative corrections to the supergravity on-shell action of the BPS black holes in some holographic models \cite{Bobev:2021qxx, Bobev:2022bjm, Cassani:2022lrk}.}

One may want to rewrite \eqref{eq:free} in terms of the superconformal $R$-symmetry chemical potential $\delta = \frac{\sigma+\tau - 1}{2}$ and flavor symmetry chemical potential $\delta^I$'s. Note that $\Delta_I$'s in \eqref{eq:free} are all in fact integer-shifted ones $\tilde\Delta_I = \Delta_I + n_I p_I$'s as explained in Section \ref{sec:sl3z}. Using the linear maps around \eqref{eq:charge-map} between charges and chemical potentials, we get
\begin{equation}\label{eq:free2}
    \begin{split}
        \log\mathcal{I}\sim-\frac{i\pi}{24\sigma\tau}&\left[k_{RRR} \,\delta_\mp^3 + 3\sum_{I=1}^{d-1}k_{RRI}\, \delta_\mp^2 \tilde\delta^I+3 \sum_{I,J=1}^{d-1}k_{RIJ} \, \delta_\mp \tilde\delta^I\tilde\delta^J + \sum_{I,J,K=1}^{d-1}k_{IJK}\tilde\delta^I\tilde\delta^J\tilde\delta^K  \right]\ ,
    \end{split}
\end{equation}
where $\delta_\mp \equiv \frac{\sigma+\tau \mp 1}{2}$,\footnote{We could have defined the index using $\delta_+ = \delta + 1$ in \eqref{eq:tr-index1} instead of $\delta = \delta_-$.} and $\tilde\delta^I \equiv \delta^I + \sum_{J=1}^dn_J p_J \xi^{JI}$ with $\sum_{I=1}^d n_I p_I =0$ for $\delta_-$ and 2 for $\delta_+$.
Various trace anomaly coefficients in the above formula are defined as
\begin{align}
    \begin{split}
        k_{RRR}&\equiv\Tr R^3=\textrm{dim}(G)+\sum_\chi \dim(\mathcal{R}_\chi)(R_\chi-1)^3 = \frac{16}{9}(5a-3c)\ ,\\
        k_{RRI}&\equiv\Tr R^2f_I=\sum_\chi\dim(\mathcal{R}_\chi)(R_\chi-1)^2f_{\chi,I}\ ,\\
        k_{RIJ}&\equiv\Tr Rf_I f_J=\sum_\chi\dim(\mathcal{R}_\chi)(R_\chi-1)f_{\chi,I}f_{\chi,J}\ ,\\
        k_{IJK}&\equiv \Tr f_If_Jf_K=\sum_\chi\dim(\mathcal{R}_\chi)f_{\chi,I}f_{\chi,J}f_{\chi,K}\ .
    \end{split}
\end{align}
Here, $R_\chi$ and $f_{\chi,I}$'s are the superconformal $R$ and flavor charges of (the scalar in) the chiral multiplet $\Phi_\chi$, and $a,c$ are the central charges or conformal anomalies of the theory.

If we turn off all the shifted flavor chemical potential $\tilde\delta^I$'s in \eqref{eq:free2}; \textit{e.g.} $\delta^I=n_I=0$ for the case with $\delta_-$, we obtain a universal large $N$ formula for the index of 4d $\CN=1$ SCFTs given by
\begin{align}\label{eq:universal}
    \begin{split}
        \log\mathcal{I}(\sigma,\tau)&\sim- \Tr R^3\; \frac{i\pi \delta_\mp^3}{24\sigma\tau} = -(5a-3c) \; \frac{2i\pi\delta_\mp^3}{27\sigma\tau}\ ,
    \end{split}
\end{align}
where $\delta_\mp = \frac{\sigma+\tau \mp 1}{2}$. One nice fact in this unrefined case is that if we take the collinear limit $\textrm{Im} (\frac{\sigma}{\tau}) \to 0$, the conditions \eqref{eq:chem-con1}, \eqref{eq:chem-con2} are trivially satisfied.\footnote{Here, we assumed $0\leq R_\chi \leq 2$ for any scalar field in the theory. For the case with $R_\chi <0$ or $R_\chi>2$, we can appropriately turn on the flavor chemical potentials so that the effective $R$-charge dressed by the flavor charges become $0 \leq \tilde R_\chi \leq 2$. Then, we can turn off the flavor chemical potentials in the final results since the index is smooth in that limit. Similar things happen for the Cardy formula \cite{Kim:2019yrz}.} Namely, the above large $N$ formula works for arbitrary collinear $\sigma,\tau$ when the shifted flavor chemical potentials are all turned off. For holographic theories, we have $a,\,c =O(N^2)$ and $a-c = O(N^1)$ or $O(N^0)$. 
Then, the above formula accounts for the Bekenstein-Hawking entropy of the universal BPS black holes in AdS$_5$ \cite{Gutowski:2004ez,Gutowski:2004yv,Chong:2005hr,Kunduri:2006ek}, arising as a solution of 5d minimal gauged supergravity whose Newton constant $G_N$ and gauge coupling $g$ are related to $c$ as $\frac{\pi^2}{g^3G_N} = c$ at $O(N^2)$.

\subsection{More general saddles}\label{sec:more}
In this subsection, we shall construct two kinds of new large $N$ saddle points generalizing the parallelogram ansatz \eqref{eqn: parallelogram ansatz (2)}, \eqref{eqn: parallelogram ansatz (3)}. The first one is $(\sigma_{r_1},\tau_{r_2})$-saddles whose edge vectors of the parallelogram ansatz \eqref{eqn: parallelogram ansatz (2)}, \eqref{eqn: parallelogram ansatz (3)} are now given by \cite{Choi:2021rxi}
\begin{equation}
    (\sigma_{r_1},\tau_{r_2})\equiv(\sigma+r_1,\tau+r_2)\ ,
\end{equation}
where $r_1, r_2 \in \mathbb{Z}$. Since $\Gamma(z;\sigma,\tau) = \Gamma(z;\sigma+1,\tau)= \Gamma(z;\sigma,\tau+1)$, one can do the very same analysis in the former subsection with $(\sigma_{r_1},\tau_{r_2})$. Then, we again perform appropriate shift of $\Delta_I$'s such that shifted chemical potentials $\Delta_I^{(r_1,r_2)},\; \sigma_{r_1}, \;\tau_{r_2}$ satisfy \eqref{eq:constraint}, \textit{i.e.}
\begin{equation}
    \sum_{I=1}^d\Delta_I^{(r_1,r_2)}-\sigma_{r_1}-\tau_{r_2}=\mp1\ .
\end{equation}
For given $\Delta_I, \sigma, \tau$, there may or may not exist the possible shift of $\Delta_I$'s satisfying the above condition and \eqref{eq:chem-con1}, \eqref{eq:chem-con2} depending on the period $p_I$'s and $(r_1, r_2)$. In principle, allowed $(r_1,r_2)$ can be determined for given $\Delta_I, \sigma, \tau$, but in general, it depends on the explicit details of the theory, which will not be discussed here. However, for the simplest case when $\sigma=\tau$ and all the flavor chemical potentials are turned off, one can show that only the cases with $r_1=r_2$ are allowed \cite{Choi:2021rxi,Choi:2022asl}.
When the above condition and \eqref{eq:chem-con1}, \eqref{eq:chem-con2} are satisfied, the $(\sigma_{r_1},\tau_{r_2})$-saddles contribute to the index in the large $N$ limit as
\begin{equation}
    \log \mathcal{I} \sim -\frac{i\pi}{24\sigma_{r_1}\tau_{r_2}}\sum_{I,J,K=1}^{d}\Tr( Q^IQ^JQ^K)\Delta_I^{(r_1,r_2)}\Delta_J^{(r_1,r_2)}\Delta_K^{(r_1,r_2)}\ .
\end{equation}
One can easily show that after the Legendre transformation, their leading large $N$ entropies are all the same \cite{Choi:2021rxi}. In dual AdS$_5$ gravity side, these saddles will correspond to multiple Euclidean solutions coming from the same Lorentzian BPS black hole solution after compactifying the temporal circle, which are explicitly constructed in asymptotically AdS$_5 \times S^5$ case \cite{Aharony:2021zkr}.

Next, we construct multi-cut saddle points in the collinear case $\textrm{Im}\left(\frac{\sigma}{\tau}\right)= 0$.\footnote{There is an issue about multi-cut saddles for non-collinear $\sigma,\tau$. For details, we refer to \cite{Choi:2021rxi}.} Our $K$-cut ansatz roughly takes the following form in the large $N$ continuum limit \cite{Choi:2021rxi}:
\begin{equation}\label{eqn: K-cut saddles}
    u_A^{(a)}(x,y)=\frac{A}{K}+\sigma x+\tau y\ ,
\end{equation}
where $A=0,1,\cdots,K-1$ labels $K$ clusters of evenly distributed eigenvalues forming the $K$-cuts. Domain of $(x,y)$ should be properly given according to the gauge group at each node as \eqref{eqn: parallelogram ansatz (2)}, \eqref{eqn: parallelogram ansatz (3)}.
Although these cuts are linear (not areal), we again adopt the $2$-parameter labeling of eigenvalues with uniform $2$-dimensional distributions. One can understand it as the collinear limit $\textrm{Im}\left(\frac{\sigma}{\tau}\right) \to 0$ of the areal distribution. Each cut contains evenly distributed $\frac{N}{K}$ eigenvalues so the density function is given by 
\begin{equation}
\rho_A^{(a)}(x,y) \to \left\{ \begin{array}{ll}
\frac{1}{K}\ ,\qquad \qquad & SU(N_a) \\
\frac{2}{K}\ ,\qquad \qquad & SO(2N_a),\; SO(2N_a+1), \;Sp(N_a)
\end{array} \right.\ .
\end{equation}
Now, let us denote the index \eqref{eqn: matrix integral-index} schematically as $\mathcal{I} = \frac{\kappa^{N}}{|W_G|}\oint d^Nu \; e^{-V(u)}$. 
Each chiral multiplet $\Phi_\chi$ contributes to the force acting on a particular eigenvalue $u_A^{(a)} (x_i,y_i)$ as
\begin{equation}
    \sum_{j}\sum_{A=0}^{K-1} \frac{\partial}{\partial u_i^{(a)}}\log \Gamma\left(u_{ij}^{(ab)} + \frac{A}{K}+ Q_\chi \cdot \Delta;\sigma,\tau\right)\ ,
\end{equation}
where $u_{ij}^{(ab)}$ denotes one of \eqref{eq:uab} and $u_i^{(a)}$ takes the value from the parallelogram ansatz \eqref{eqn: parallelogram ansatz (2)}, \eqref{eqn: parallelogram ansatz (3)} as before. Once again, the vector multiplet contribution can be understood with overall minus sign, and identifying $u_{ij}^{(aa)} = \alpha_a(u)$ and $Q_\chi=(0,0,\cdots,0)$. Using the following identity
\begin{equation}
    \sum_{A=0}^{K-1}\log\Gamma\left(z+\frac{A}{K};\sigma,\tau\right)=\log\Gamma(Kz;K\sigma,K\tau)\ ,
\end{equation}
above formula is simplified as
\begin{equation}
\begin{aligned}
    & K \sum_{j} \frac{\partial}{\partial {u_i^{(a)}}'} \log \Gamma\left({u_{ij}^{(ab)}}' +  Q_\chi \cdot \Delta';\sigma',\tau'\right)\ ,
    \end{aligned}
\end{equation}
where $\Delta_I' \equiv K \Delta_I + n_I^{(K)}p_I$, $\sigma' \equiv K\sigma$, $\tau' \equiv K\tau$, and ${u_i^{(a)}}' \equiv K u_i^{(a)}$. Here, $n_I^{(K)} \in \mathbb{Z}$ are appropriate integer shifts of $K\Delta_I$'s so that
\begin{align}
    \sum_{I=1}^d\Delta_I^\prime-\sigma^\prime-\tau^\prime=\mp1\ .
\end{align}
Such shifts may or may not exist depending on $p_I$'s.
The above force takes ($K$ times) the precisely same form with that of the 1-cut solution with $\Delta_I', \sigma', \tau'$ before the $SL(3,\mathbb{Z})$ transformation. Therefore, our $K$-cut ansatz \eqref{eqn: K-cut saddles} indeed represents large $N$ saddles generalizing 1-cut solution \eqref{eqn: parallelogram ansatz (2)}, \eqref{eqn: parallelogram ansatz (3)} if $\Delta_I', \sigma', \tau'$ satisfy the above condition and \eqref{eq:chem-con1}, \eqref{eq:chem-con2}.
Then, it is straightforward to show that each chiral multiplet $\Phi_\chi$ contributes to $V(u)$ as
\begin{equation}
\begin{aligned}
    &v(u_{ij}^{(ab)}) = - K \sum_{i,j} \log \Gamma\left(Ku_{ij}^{(ab)} +  Q_\chi \cdot (K\Delta);K\sigma,K\tau\right) \\
    &\xrightarrow{N\to \infty} - \frac{N_a N_b}{K} \int_{-1/2}^{1/2}\int_{-1/2}^{1/2}\int_{-1/2}^{1/2}\int_{-1/2}^{1/2}dx_1dy_1 dx_2dy_2 \log \Gamma\left(u'_{12} +  Q_\chi \cdot \Delta';\sigma',\tau'\right) \ ,
    \end{aligned}
\end{equation}
where $u'_{12} \equiv \sigma' x_{12} + \tau' y_{12}$. Here, we used the periodicity $\Gamma(z+1;\sigma,\tau) = \Gamma(z;\sigma,\tau)$ to reduce the double sum over cuts to the single sum. Vector multiplet contribution can be treated as before.
The above potential is the same as $\frac{1}{K}$ times that of the 1-cut solution with primed variables.
As a result, the $K$-cut solution contributes to the large $N$ free energy as
\begin{equation}
    \log\mathcal{I}\sim -\frac{i\pi}{24K\sigma^\prime\tau^\prime}\sum_{I,J,K=1}^d\Tr( Q^IQ^JQ^K)\Delta_I^\prime\Delta_J^\prime \Delta_K^\prime\ ,
\end{equation}
whose leading large $N$ entropy, after the Legendre transformation, is $\frac{1}{K}$ times that of the 1-cut solution \cite{Choi:2021rxi}. In dual AdS$_5$ gravity side, these saddles will be dual to $\mathbb{Z}_K$ orbifolds of the Euclidean BPS black hole solutions dual to the 1-cut solution, which are explicitly constructed in asymptotically AdS$_5 \times S^5$ case \cite{Aharony:2021zkr}.

One can also combine the above two saddles to make more general saddles. Then, their contributions to the index will be
\begin{equation}
    \log\mathcal{I}\sim -\frac{i\pi}{24K\sigma'_{r_1}\tau'_{r_2}}\sum_{I,J,K=1}^d\Tr( Q^IQ^JQ^K){\Delta_I'}^{(r_1,r_2)}{\Delta_J'}^{(r_1,r_2)}{\Delta_K'}^{(r_1,r_2)}\ ,
\end{equation}
where $\sigma'_{r_1} \equiv K\sigma+r_1$, $\tau'_{r_2} \equiv K\tau + r_2$, and ${\Delta_I'}^{(r_1,r_2)} \equiv K\Delta_I + n_I^{(K,r_1,r_2)}p_I$ satisfy
\begin{equation}
    \sum_{I=1}^d{\Delta_I'}^{(r_1,r_2)}-{\sigma'}_{r_1}-{\tau'}_{r_2}=\mp1\ ,
\end{equation}
and \eqref{eq:chem-con1} or \eqref{eq:chem-con2} for $n_I^{(K,r_1,r_2)} \in \mathbb{Z}$. In principle, we should sum over contributions from all of these saddles labeled by $(K,r_1,r_2, \mp)$ as the large $N$ approximation of the index. However, it is not known whether this sum is convergent. Also, it is not known if all the contributions should be summed with equal weights within the framework of the Picard-Lefschetz theory. 
Here, we just state that we have a one-to-one correspondence between these kinds of large $N$ saddles for $\CN=4$ $U(N)$ SYM \cite{Choi:2021rxi} and (orbifolded) Euclidean black hole solutions in AdS$_5 \times S^5$ \cite{Aharony:2021zkr}. Further, note that they are also in one-to-one correspondence with the large $N$ limit of the Hong-Liu solutions \cite{Hong:2018viz} of the Bethe Ansatz Equation for the index of $\CN=4$ $SU(N)$ SYM \cite{Benini:2018ywd}.

Let us make a final comment before closing this section. In \cite{Choi:2021rxi}, multi-cut saddle points with unequal filling fractions were found in $\CN=4$ $U(N)$ SYM. Here, unequal filling fractions mean that the number of eigenvalues occupying each parallelogram is different. Neither corresponding solutions in AdS$_5$ gravity nor Bethe roots of the index are known. It would be interesting to construct this kind of large $N$ solutions for $\mathcal{N}=1$ gauge theories, generalizing the multi-cut saddle points in this subsection. Since such solutions highly depend on the details of the theory, we shall not discuss them here.

\section{Examples}\label{sec:examples}
In this section, we invoke the large $N$ analysis in Section \ref{section1} against several concrete holographic models.
For brevity, we will omit all the modular parameters of the elliptic gamma functions in the following examples, regardless of whether they have been $SL(3,\mathbb{Z})$ transformed or not.

\subsection{Klebanov-Witten theory}

The Klebanov-Witten theory \cite{Klebanov:1998hh} is a 4d $\mathcal{N}=1$ supersymmetric gauge theory, which arises as a low-energy description of a stack of $N$ D3-branes sitting at a conifold singularity. This theory is dual to type IIB string theory on AdS$_5 \times T^{1,1}$.

The theory has $SU(N)\times SU(N)$ gauge group with two pairs of bifundamental chiral multiplets $A_{1,2},B_{1,2}$ transforming in the $(\mathbf{N},\overline{\mathbf{N}})$ and $(\overline{\mathbf{N}},\mathbf{N})$ representations. The quartic superpotential is given by
\begin{equation}
    \mathcal{W}=\tr\big(A_1B_1A_2B_2-A_1B_2A_2B_1)\ .
\end{equation}
Apart from the conformal symmetry, the bosonic global symmetry of the theory involves the $R$-symmetry, two $SU(2)$ flavor symmetries rotating the $A_{1,2}$ and $B_{1,2}$ doublets, and a $U(1)_B$ baryonic symmetry as follows:
\begin{equation}
    U(1)_{R}\times U(1)_B \times SU(2)_{F_1}\times SU(2)_{F_2}\ .
\end{equation}
The charge of each multiplet under the maximal torus of the global symmetry is presented in Table~\ref{tab: Charge assignment global symmetries for KW theory}. In general, the $R$-charge $R_0$, determined by the anomaly-free condition and $R[\mathcal{W}]=2$, must be considered as candidate $R$-charge. However, for the Klebanov-Witten theory, it becomes the true $R$-charge as $U(1)_B$ is enhanced to $SU(2)_B$ at IR. 

\begin{table}[t]
\small
    \centering
    \begin{tabular}{|c|c|c|c|c||c|c|c|c|}
    \hline
     & $U(1)_{F_1}$ & $U(1)_{F_2}$ & $U(1)_B$ & $U(1)_{R_0}$ & $U(1)_{1}$ & $U(1)_{2}$ & $U(1)_{3}$ & $U(1)_{4}$\\ \hline
     $A_1$ & $1$ & $0$ & $1$ & $1/2$ & $1$ & $0$ & $0$ & $0$\\\hline 
     $A_2$ & $-1$ & $0$ & $1$ & $1/2$ & $0$ & $1$ & $0$ & $0$\\ \hline
     $B_1$ & $0$ & $1$ & $-1$ & $1/2$ & $0$ & $0$ & $1$ & $0$\\ \hline
     $B_2$ & $0$ & $-1$ & $-1$ & $1/2$ & $0$ & $0$ & $0$ & $1$\\ \hline
    \end{tabular}
    \caption{Charge assignment of each chiral multiplet in the Klebanov-Witten theory}
    \label{tab: Charge assignment global symmetries for KW theory}
\end{table}

For clarification, let us begin with the usual definition of the superconformal index with the $(-1)^F$ insertion
\begin{equation}\label{eqn: tr-index of KW (1)}
    \mathcal{I}(\xi,\sigma,\tau)=\Tr \left[(-1)^Fe^{2\pi i\sigma \left(J_1+\frac{R}{2}\right)}e^{2\pi i\tau\left(J_2+\frac{R}{2}\right)}e^{2\pi i\left(\xi_{F_1}Q_{F_1}+\xi_{F_2}Q_{F_2}+\xi_{B}Q_{B}\right)}\right]\ .
\end{equation}
As explained thoroughly in section \ref{section1}, it is convenient to redefine the flavor chemical potentials in terms of
\begin{equation}
    \begin{split}
        \Delta_1&=\xi_{F_1}+\xi_{B}+\frac{1}{4}(\tau+\sigma \mp 1)\ , \qquad\Delta_2=-\xi_{F_1}+\xi_{B}+\frac{1}{4}(\tau+\sigma \mp 1)\ ,\\
        \Delta_3&=\xi_{F_2}-\xi_{B}+\frac{1}{4}(\tau+\sigma \mp 1)\ , \qquad\Delta_4=-\xi_{F_2}-\xi_{B}+\frac{1}{4}(\tau+\sigma \mp 1)\ ,
    \end{split}
\end{equation}
which satisfy the following relation:
\begin{equation}\label{eqn: two relevant branches of KW theory}
    \Delta_1+\Delta_2+\Delta_3+\Delta_4=\tau+\sigma\mp1\ .
\end{equation}
Then, the index takes a more transparent form of \eqref{eqn: tr-index}:
\begin{equation}\label{eqn: tr-index of KW (2)}
    \mathcal{I}(\Delta,p,q)=\Tr \left[e^{2\pi i\sigma J_1}e^{2\pi i\tau J_2}\prod_{I=1}^{4}e^{2\pi i\Delta_I Q^I}\right]\ ,
\end{equation}
where new charge $Q^I$'s are integer-normalized as in Table~\ref{tab: Charge assignment global symmetries for KW theory}. Here, we have assigned chemical potential $\Delta_I$'s to each bifundamental fields. In other words, since $2Q^I$ plays the role of $R$-symmetry, there is one-to-one correspondence between bifundamental field and its $R$-symmetry, like $\mathcal{N}=4$ SYM case. Now we can write the integral representation of the index \eqref{eqn: tr-index of KW (2)} as
\begin{align}
    \begin{split}
        \mathcal{I}(\Delta,\sigma,\tau)=&\frac{\kappa^{2N}}{(N!)^2}\oint\prod_{i=1}^{N}du_i^{(1)}du_i^{(2)}\frac{1}{\prod_{i\neq j}^N\Gamma\Big(u_{ij}^{(11)}\Big)\Gamma\Big(u_{ij}^{(22)}\Big)}\\
        &\times\prod_{i,j=1}^N\Gamma\Big(u_{ij}^{(12)}+\Delta_{1}\Big)\Gamma\Big(u_{ij}^{(12)}+\Delta_{2}\Big)\Gamma\Big(u_{ij}^{(21)}+\Delta_{3}\Big)\Gamma\Big(u_{ij}^{(21)}+\Delta_{4}\Big)\ ,
    \end{split}
\end{align}
where $u_{ij}^{(ab)}\equiv u_i^{(a)}-u_j^{(b)}$.

Using the $SL(3,\mathbb{Z})$ identity \eqref{eqn: SL(3,Z) identity (1)} assuming $\textrm{Im} \left(\frac{\sigma}{\tau}\right)>0$, the above integrand becomes
\begin{equation}
    \mathcal{I}(\Delta,\sigma,\tau)=\frac{\kappa^{2N}}{(N!)^2}\oint\prod_{i=1}^N du_i^{(1)}du_i^{(2)}e^{-i\pi\mathbf{P}_+}e^{-\sum_{i,j=1}^N\big(V_\sigma(u)+V_\tau(u)\big)}\ ,
\end{equation}
where (ignoring $i=j$ terms of the vector multiplet contribution)
\begin{align}
    V_\sigma(u)&\equiv\frac{1}{2}\sum_{I=1}^4\log\Gamma\left(-\frac{u_{ij}^{(12)}+\Delta_I+1}{\sigma}\right)-\frac{1}{2}\log\Gamma\left(-\frac{u_{ij}^{(11)}+1}{\sigma}\right)-\frac{1}{2}\log\Gamma\left(-\frac{u_{ij}^{(22)}+1}{\sigma}\right)\nonumber\\
    &\qquad+(u\rightarrow-u)\ ,\\
    V_\tau(u)&\equiv\frac{1}{2}\sum_{I=1}^4\log\Gamma\left(\frac{u_{ij}^{(12)}+\Delta_I+1}{\tau}\right)-\frac{1}{2}\log\Gamma\left(\frac{u_{ij}^{(11)}}{\tau}\right)-\frac{1}{2}\log\Gamma\left(\frac{u_{ij}^{(22)}}{\tau}\right) \nonumber\\
    &\qquad+(u\rightarrow-u)\ .\nonumber
\end{align}
For convenience, we averaged over the contributions from positive and negative roots and weights to have even function of $u$, which is true for the Klebanov-Witten theory. Then, the cubic and linear terms in $u$ of the prefactor are automatically cancelled, and hence we obtain
\begin{align}
    \begin{split}
        -i&\pi\mathbf{P}_\pm\Big|_{\eqref{eqn: two relevant branches of KW theory}}\\
        =&-\frac{i\pi}{2} \Bigg[\sum_{i,j=1}^N\sum_{I=1}^4P_\pm(u_{ij}^{(12)}+\Delta_I)-\sum_{i\neq j}^NP_\pm(u_{ij}^{(11)})-\sum_{i\neq j}^NP_\pm(u_{ij}^{(22)})\Bigg]+(u\rightarrow-u)\\
        =&-\frac{i\pi N^2(\Delta_1\Delta_2\Delta_3+\Delta_1\Delta_2\Delta_4+\Delta_1\Delta_3\Delta_4+\Delta_2\Delta_3\Delta_4)}{\sigma\tau}\\
        &-\frac{i\pi N(\sigma+\tau-\sigma\tau)}{6\sigma\tau}(\Delta_1+\Delta_2+\Delta_3+\Delta_4)\\
        &+\frac{i\pi(\sigma+\tau\mp1)}{2\sigma\tau}\left[\sum_{i,j=1}^N2(u_{ij}^{(12)})^2-\sum_{i\neq j}^N(u_{ij}^{(11)})^2-\sum_{i\neq j}^N(u_{ij}^{(22)})^2\right]\ .
    \end{split}
\end{align}
From the discussion in section \ref{section1}, we recognize that the $u$-dependent part can be rewritten as
\begin{align}
    \sum_{i,j=1}^N\big[2(u_{ij}^{(12)})^2-(u_{ij}^{(11)})^2-(u_{ij}^{(22)})^2\big]&=\big(2d_2(\mathbf{N})\dim(\overline{\mathbf{N}})+2\dim(\mathbf{N})d_2(\overline{\mathbf{N}})-2d_2(\textrm{Adj})\big)|u|^2\nonumber\\
    &=\left(4\cdot\frac{1}{2}\cdot N-2N\right)|u|^2=0\ , 
\end{align}
since $2Q^I-1=1$. In addition, since the overall gauge group is $SU(N)\times SU(N)$, we only need to use the first type of parallelogram ansatz \eqref{eqn: parallelogram ansatz (2)} to solve the large $N$ matrix model. In doing so, we should split the dangerous Haar-measure-like factor from each gauge node, and then use the integral identity \eqref{eqn: integral identity}. This leads to 
\begin{equation}
    \mathcal{I}(\Delta,\sigma,\tau)=\kappa^{2N}\oint\prod_{c=1}^{2}\prod_{i=1}^{N}du_i^{(c)}\prod_{i<j}^{N}\left(1-e^{\frac{2\pi iu_{ij}^{(cc)}}{\tau}}\right)e^{-i\pi\mathbf{P}_+}e^{-\sum_{i,j=1}^N\big(V_\sigma(u)+\Tilde{V}_\tau(u)\big)}\ ,
\end{equation}
where both $V_\sigma,\Tilde{V}_\tau$ and the factorized half-Haar-measure-like factor have vanishing force within the parallelogram region if the following constraints \eqref{eqn: four-chemical potential inequalities} are met:
\begin{align}\label{eqn: constraints of chemical potentials of KW}
    \Im\left(\frac{1-\tau+\Delta_I}{\sigma}\right)<0,\; \Im\left(\frac{\Delta_I}{\sigma}\right)>0,\;\Im\left(\frac{-\sigma+\Delta_I}{\tau}\right)>0,\; \Im\left(\frac{1+\Delta_I}{\tau}\right)<0\ .
\end{align}
It also follows that the potentials $V_\sigma,\Tilde{V}_\tau$ themselves vanish at the parallelogram saddle point at leading $O(N^2)$ order.
Consequently, the only terms that contribute to the free energy is the $u$-independent part of the prefactor $-i\pi\mathbf{P}_\pm$, which has the same structure as \eqref{eq:free}:
\begin{align}
    \begin{split}
        \log\mathcal{I}(\Delta,\sigma,\tau)\sim&-\frac{i\pi N^2(\Delta_1\Delta_2\Delta_3+\Delta_1\Delta_2\Delta_4+\Delta_1\Delta_3\Delta_4+\Delta_2\Delta_3\Delta_4)}{\tau \sigma}\ ,
    \end{split}
\end{align}
where we only kept terms of order $O(N^2)$. This perfectly matches the entropy function of the BPS black holes in AdS$_5 \times T^{1,1}$, which was obtained in the gravity side when $\sigma=\tau$ and only the chemical potentials for the $R$-symmetry $U(1)_R$ and baryonic symmetry $U(1)_B$ are turned on \cite{Benini:2020gjh}.

\subsection{$Y^{p,q}$-theories}

Let us consider the low energy dynamics of a stack of D3-branes probing a nontrivial toric Calabi-Yau 3-fold singularity. As a concrete example, we will study an infinite family of $\CN=1$ quiver gauge theories dual to type IIB string theory on AdS$_5 \times Y^{p,q}$, where $p>q$ are positive integers \cite{Martelli:2004wu,Benvenuti:2004dy}. $Y^{p,q}$'s are one of the well-known examples of the toric Sasaki-Einstein 5-manifolds.

The theory has $2p$ $SU(N)$ gauge nodes with $4p+2q$ bifundamental chiral multiplets categorized by four types: $p$ fields of type $U_{1,2}$, $q$ fields of type $V_{1,2}$, $p-q$ fields of type $Z$, and $p+q$ fields of type $Y$. The superpotential is then made by summing over all possible cubic and quartic gauge invariant operators using $U_\alpha,V_\alpha,Y$ and $Z$ as follows:
\begin{equation}\label{eqn: superpotential Ypq}
    \mathcal{W}=\sum_k\epsilon^{\alpha\beta}\tr\big(U_\alpha^{(k)} V_\beta^{(k)} Y^{(2k+2)}+V_\alpha^{(k)} U_\beta^{(k+1)}Y^{(2k+3)}\big)+\sum_k\epsilon^{\alpha\beta}\tr\big(Z^{(k)} U_\alpha^{(k+1)}Y^{(2k+3)}U_\beta^{(k)}\big)\ .
\end{equation}
In addition to the conformal symmetry, the global symmetry of the theory is given as
\begin{equation}
    U(1)_R\times U(1)_B\times U(1)_{F_1}\times SU(2)_{F_2}\ ,
\end{equation}
whose charge assignments to each bifundamental field is summarized in Table~\ref{tab: Charge assignment global symmetries for Ypq theory}, where we again redefined new integer charges $Q^I$'s appropriately. 
Then, the corresponding chemical potentials $\Delta_{I=1,\cdots,6}$'s are assigned to each type of bifundamental field, but they are not entirely independent:
\begin{equation}\label{eqn: two relevant branches of Ypq}
    \Delta_5=\Delta_2+\Delta_3\ ,\quad \Delta_6=\Delta_3+\Delta_4\ ,\quad\sum_{I=1}^4\Delta_{I}-\sigma-\tau=\mp1\ .
\end{equation}
The integral representation of the index for $Y^{p,q}$-theories can be written as
\begin{equation}
    \mathcal{I}_{p,q}(\Delta,\sigma,\tau)=\frac{\kappa^{2pN}}{(N!)^{2p}}\oint\prod_{c=1}^{2p}\prod_{i=1}^{N}du_i^{(c)}\frac{\prod_{a\rightarrow b}\prod_{i,j=1}^{N}\Gamma\Big(u_{ij}^{(ab)}+\Delta_{I}\Big)}{\prod_{i\neq j}^N\Gamma\Big(u_{ij}^{(cc)}\Big)}\ ,
\end{equation}
where $u_{ij}^{(ab)}\equiv u_i^{(a)}-u_j^{(b)}$ and the symbol $\prod_{a\rightarrow b}$ denotes the product over all bifundamental chiral multiplet contributions associated with arrows that start from any node $a$ and reach to any node $b$. Note that the chemical potentials $\Delta_I$'s depend implicitly on the gauge node index $a,b$ in the above expression.

\begin{table}[t]
\small
\centering
    \begin{tabular}{|c|c||c|c|c|c||c|c|c|c|}
    \hline
     & Multiplicity & $U(1)_{F_1}$ & $U(1)_{F_2}$ & $U(1)_B$ & $U(1)_{R_0}$ & $U(1)_1$ & $U(1)_2$ & $U(1)_3$ & $U(1)_4$ \\ \hline
     $Y$ & $p+q$ & $1$ & $0$ & $0$ & $\frac{1}{2}$ & $1$ & $0$ & $0$ & $0$ \\\hline
     $U_1$ & $p$ & $0$ & $1$ & $0$ & $\frac{1}{2}$ & $0$ & $1$ & $0$ & $0$ \\\hline
     $Z$ & $p-q$ & $0$ & $0$ & $1$ & $\frac{1}{2}$ & $0$ & $0$ & $1$ & $0$ \\\hline
     $U_2$ & $p$ & $-1$ & $-1$ & $-1$ & $\frac{1}{2}$ & $0$ & $0$ & $0$ & $1$ \\\hline
     $V_1$ & $q$ & $0$ & $1$ & $1$ & $1$ & $0$ & $1$ & $1$ & $0$\\\hline
     $V_2$ & $q$ & $-1$ & $-1$ & $0$ & $1$ & $0$ & $0$ & $1$ & $1$ \\ \hline
    \end{tabular}
    \caption{Charge assignment of each chiral multiplet in the $Y^{p,q}$-theories}
    \label{tab: Charge assignment global symmetries for Ypq theory}
\end{table}

Following the same steps, applying the $SL(3,\mathbb{Z})$ identity \eqref{eqn: SL(3,Z) identity (1)} assuming $\textrm{Im} \left(\frac{\sigma}{\tau}\right)>0$, splitting the dangerous Haar-measure-like factor from each gauge node, and using the integral identity \eqref{eqn: integral identity}, we obtain
\begin{align}\label{eqn: matrix model of Ypq}
    \begin{split}
        \mathcal{I}_{p,q}&(\Delta,\sigma,\tau)\\
        &=\kappa^{2pN}\oint\prod_{c=1}^{2p}\prod_{i=1}^{N}du_i^{(c)}\prod_{i<j}^{N}\left(1-e^{\frac{2\pi iu_{ij}^{(cc)}}{\tau}}\right)e^{-i\pi\mathbf{P}_+}e^{-\sum_{i,j=1}^N\big(V_\sigma(u)+\Tilde{V}_\tau(u)\big)}\ ,
    \end{split}
\end{align}
where (ignoring $i=j$ terms of the vector multiplet contribution)
\begin{align}
    \begin{split}
        -V_\sigma(u)&\equiv\frac{1}{2}\sum_{a\rightarrow b}\log\Gamma\bigg(-\frac{u_{ij}^{(ab)}+\Delta_{I_{ab}}+1}{\sigma}\bigg)-\frac{1}{2}\sum_{c=1}^{2p}\log\Gamma\bigg(-\frac{u_{ij}^{(c)}+1}{\sigma}\bigg)\ ,\\
        -\Tilde{V}_\tau(u)&\equiv\frac{1}{2}\sum_{a\rightarrow b}\log\Gamma\bigg(\frac{u_{ij}^{(ab)}+\Delta_{I_{ab}}}{\tau}\bigg)-\frac{1}{2}\sum_{c=1}^{2p}\log\Gamma\bigg(\frac{u_{ij}^{(c)}}{\tau}\bigg)\ ,
    \end{split}
\end{align}
and the prefactor is given by
\begin{equation}
    \begin{split}
        -i\pi\mathbf{P}_\pm&=-i\pi\sum_{i,j=1}^N\bigg[\sum_{a=1}^p P_\pm(\Delta_2+u_{ij}^{(2a,2a+1)})+\sum_{a=1}^p P_\pm(\Delta_4+u_{ij}^{(2a,2a+1)})\\
        &\qquad+\sum_{a=1}^qP_\pm(\Delta_2+\Delta_3+u_{ij}^{(2a-1,2a)})+\sum_{a=1}^q P_\pm(\Delta_3+\Delta_4+u_{ij}^{(2a-1,2a)})\\
        &\qquad+\sum_{a=1}^{2q}P_\pm(\Delta_1+u_{ij}^{(a+1,a-1)})+\sum_{a=1}^{p-q}P_\pm(\Delta_1+u_{ij}^{(2a+2q+1,2a+2q-2)})\\
        &\qquad+\sum_{a=1}^{p-q}P_\pm(\Delta_3+u_{ij}^{(2a+2q-1,2a+2q)})\bigg]-\sum_{i\neq j}^N\sum_{a=1}^{2p}P_\pm(u_{ij}^{(a,a)})\ .
    \end{split}
\end{equation}
Substituting the definition of the cubic polynomials \eqref{eqn: Cubic polynomials}, it can be reorganized as
\begin{align}\label{eq:pre-ypq}
    -i\pi&\mathbf{P}_\pm\Big|_{\eqref{eqn: two relevant branches of Ypq}}\nonumber\\
    =&-\frac{\pi iN^2\big[p\Delta_1\Delta_2\Delta_3+(p+q)\Delta_1\Delta_2\Delta_4+p\Delta_1\Delta_3\Delta_4+(p-q)\Delta_2\Delta_3\Delta_4\big]}{\sigma\tau}\nonumber\\
    &-\frac{i\pi Np(\sigma+\tau-\sigma\tau)}{6\sigma\tau}(\Delta_1+\Delta_2+\Delta_3+\Delta_4)\nonumber\\
    &-\frac{i\pi}{2\sigma\tau}\Delta_1\sum_{i,j}^N\bigg[\sum_{a=1}^{2q}\big(u_{ij}^{(a+1,a-1)}\big)^2+\sum_{a=1}^{p-q}\big(u_{ij}^{(2a+2q+1,2a+2q-2)}\big)^2+\sum_{a=1}^{2p}\big(u_{ij}^{(a,a)}\big)^2\nonumber\\
    &\qquad\qquad-2\sum_{a=1}^p\big(u_{ij}^{(2a,2a+1)}\big)^2-2\sum_{a=1}^q\big(u_{ij}^{(2a-1,2a)}\big)^2-\sum_{a=1}^{p-q}\big(u_{ij}^{(2a+2q-1,2a+2q)}\big)^2\bigg]\nonumber\\
    &-\frac{i\pi}{2\sigma\tau}\Delta_2\sum_{i,j}^N\bigg[\sum_{a=1}^{p}\big(u_{ij}^{(2a,2a+1)}\big)^2+\sum_{a=1}^{2p}\big(u_{ij}^{(a,a)}\big)^2-\sum_{a=1}^{p}\big(u_{ij}^{(2a,2a+1)}\big)^2\\
    &\qquad\qquad-\sum_{a=1}^{2q}\big(u_{ij}^{(a+1,a-1)}\big)^2-\sum_{a=1}^{p-q}\big(u_{ij}^{(2a+2q+1,2a+2q-2)}\big)^2-\sum_{a=1}^{p-q}\big(u_{ij}^{(2a+2q-1,2a+2q)}\big)^2\bigg]\nonumber\\
    &-\frac{i\pi}{2\sigma\tau}\Delta_3\sum_{i,j}^N\bigg[2\sum_{a=1}^{q}\big(u_{ij}^{(2a-1,2a)}\big)^2+\sum_{a=1}^{p-q}\big(u_{ij}^{(2a+2q-1,2a+2q)}\big)^2+\sum_{a=1}^{2p}\big(u_{ij}^{(a,a)}\big)^2\nonumber\\
    &\qquad\qquad-2\sum_{a=1}^{p}\big(u_{ij}^{(2a,2a+1)}\big)^2-\sum_{a=1}^{2q}\big(u_{ij}^{(a+1,a-1)}\big)^2-\sum_{a=1}^{p-q}\big(u_{ij}^{(2a+2q+1,2a+2q-2)}\big)^2\bigg]\nonumber\\
    &-\frac{i\pi}{2\sigma\tau}\Delta_4\sum_{i,j}^N\bigg[\sum_{a=1}^{p}\big(u_{ij}^{(2a,2a+1)}\big)^2+\sum_{a=1}^{2p}\big(u_{ij}^{(a,a)}\big)^2-\sum_{a=1}^{p}\big(u_{ij}^{(2a,2a+1)}\big)^2\nonumber\\
    &\qquad\qquad-\sum_{a=1}^{2q}\big(u_{ij}^{(a+1,a-1)}\big)^2-\sum_{a=1}^{p-q}\big(u_{ij}^{(2a+2q+1,2a+2q-2)}\big)^2-\sum_{a=1}^{p-q}\big(u_{ij}^{(2a+2q-1,2a+2q)}\big)^2\bigg]\nonumber\ .
\end{align}
The linear terms in $u$ vanish due to the third identity in \eqref{eqn: group identities}. The cubic terms proportional to $d_3(\mathcal{R})$ also vanish since the matter fields are in the real representations under $G=\prod_{a=1}^{2p}SU(N_a)$, balancing fundamental and anti-fundamental contributions. On the other hand, the coefficients of quadratic terms, for instance $\Delta_1|u|^2$, can be recast as
\begin{align}
    \big(2q+(p-q)-2p&-2q-(p-q)\big)\cdot \big(d_2(\mathbf{N})\dim(\overline{\mathbf{N}})+\dim(\mathbf{N})d_2(\overline{\mathbf{N}})\big)+2p\cdot d_2(Adj)\nonumber\\
    &=-2p\cdot\left(2\cdot\frac{1}{2}\cdot N\right)+2p\cdot N=0\ .
\end{align}
Finally, assuming the constraints \eqref{eqn: two relevant branches of Ypq} and
\begin{align}
    \Im\left(\frac{1-\tau+\Delta_I}{\sigma}\right)<0,\; \Im\left(\frac{\Delta_I}{\sigma}\right)>0,\;\Im\left(\frac{-\sigma+\Delta_I}{\tau}\right)>0,\; \Im\left(\frac{1+\Delta_I}{\tau}\right)<0\ ,
\end{align}
we find that the parallelogram ansatz \eqref{eqn: parallelogram ansatz (2)} solves the large $N$ matrix model \eqref{eqn: matrix model of Ypq} with vanishing potentials. Therefore, the constant part of the prefactor \eqref{eq:pre-ypq} gives the large $N$ free energy as
\begin{align}
    \begin{split}
        \log\mathcal{I}(\Delta,\sigma,\tau)\sim-\frac{\pi iN^2\big[p\Delta_1\Delta_2\Delta_3+(p+q)\Delta_1\Delta_2\Delta_4+p\Delta_1\Delta_3\Delta_4+(p-q)\Delta_2\Delta_3\Delta_4\big]}{\sigma\tau}\ ,
    \end{split}
\end{align}
in accordance with the general form \eqref{eq:free}. This result is consistent with \cite{Benini:2020gjh, Amariti:2019mgp, Lanir:2019abx,  GonzalezLezcano:2019nca,Cabo-Bizet:2020nkr}.

\subsection{$\mathcal{N}=4$ SYM with $G=SO/Sp$}
As the last set of examples, we consider the $\CN=4$ SYM theories with $SO(N)$ or $Sp(N/2)$\footnote{Our convention is $C_n = Sp(n)$ so that $Sp(n)$ has rank $n$.} (for even $N$) gauge group, which lives on $N$ D3-branes probing an parallel O3$^\pm$-plane respectively. These theories are dual to type IIB string theory on AdS$_5 \times \mathbb{RP}^5$ where $\mathbb{RP}^5 = S^5/\mathbb{Z}_2$.
There exists an additional NS-NS $B$-field for the $Sp(N/2)$ case compared to the $SO(N)$ case \cite{Witten:1998xy}.

The index admits following matrix integral representation:
\begin{itemize}
    \item $Sp(N)$:
    \begin{equation}\label{eqn: integral representations of Sp(N)}
        \mathcal{I}=\frac{\kappa^N}{2^{N}N!}\prod_{I=1}^3\Gamma(\Delta_I)^N\oint\prod_{i=1}^Ndu_i\frac{\prod_{I=1}^3\prod_{i<j}^N\Gamma(\pm u_i\pm u_j+\Delta_I)\prod_{i=1}^N\Gamma(\pm2u_i+\Delta_I)}{\prod_{i<j}^N\Gamma(\pm u_i\pm u_j)\prod_{i=1}^N\Gamma(\pm2u_i)}\ 
    \end{equation}

    \item $SO(2N+1)$:
    \begin{equation}\label{eqn: integral representations of SO(2N+1)}
        \mathcal{I}=\frac{\kappa^N}{2^{N}N!}\prod_{I=1}^3\Gamma(\Delta_I)^N\oint\prod_{i=1}^Ndu_i\frac{\prod_{I=1}^3\prod_{i<j}^N\Gamma(\pm u_i\pm u_j+\Delta_I)\prod_{i=1}^N\Gamma(\pm u_i+\Delta_I)}{\prod_{i<j}^N\Gamma(\pm u_i \pm u_j)\prod_{i=1}^N\Gamma(\pm u_i)}\ 
    \end{equation}

    \item $SO(2N)$:
    \begin{equation}\label{eqn: integral representations of SO(2N)}
        \mathcal{I}=\frac{\kappa^N}{2^{N-1}N!}\prod_{I=1}^3\Gamma(\Delta_I)^N\oint\prod_{i=1}^Ndu_i\frac{\prod_{I=1}^3\prod_{i<j}^N\Gamma(\pm u_i\pm u_j+\Delta_I)}{\prod_{i<j}^N\Gamma(\pm u_i \pm u_j)}\ 
    \end{equation}
\end{itemize}
where the chemical potentials must satisfy
\begin{equation}\label{eqn: two relevant branches of N=4 with other G}
    \Delta_1+\Delta_2+\Delta_3-\sigma-\tau=\mp1\ ,
\end{equation}
and we adopted the following shorthand notations
\begin{equation}
    \begin{split}
        f(a\pm b)&\equiv f(a+b)f(a-b)\\
        f(\pm a\pm b)&\equiv f(a+b)f(a-b)f(-a+b)f(-a-b)\ .
    \end{split}
\end{equation}

Applying the $SL(3,\mathbb{Z})$ identity \eqref{eqn: SL(3,Z) identity (1)} assuming $\textrm{Im} \left(\frac{\sigma}{\tau}\right)>0$, the indices \eqref{eqn: integral representations of Sp(N)}, \eqref{eqn: integral representations of SO(2N+1)} and \eqref{eqn: integral representations of SO(2N)} will take the following form:
\begin{equation}\label{eq:n=4sl3z}
    \mathcal{I}(\Delta,\sigma,\tau)=\frac{\kappa^N}{|W_G|}\oint\prod_{i=1}^Ndu_i\;e^{-i\pi\mathbf{P}_+}e^{-\sum_{i<j}\big(V_\sigma(u)+V_\tau(u)\big)}\ ,
\end{equation}
where the prefactor $-i\pi\mathbf{P}_\pm$ in each case is naturally independent of $u$ like $SU(N)$ case \cite{Choi:2021rxi}:
\begin{equation}
    -i\pi\mathbf{P}_\pm\big|_{\eqref{eqn: two relevant branches of N=4 with other G}}=\begin{cases}
        -\frac{i\pi N(2N+1)\Delta_1\Delta_2\Delta_3}{\sigma\tau}\qquad&:Sp(N),SO(2N+1)\\
        -\frac{i\pi N(2N-1)\Delta_1\Delta_2\Delta_3}{\sigma\tau}\qquad&:SO(2N)
    \end{cases}\ .
\end{equation}
The remaining potentials are given by
\begin{align}
    \begin{split}
        -V_\sigma(u)&\sim\sum_{I=1}^3\log\Gamma\left(-\frac{\pm u_i\pm u_j+\Delta_I+1}{\sigma}\right)-\log\Gamma\left(-\frac{\pm u_i\pm u_j+1}{\sigma}\right)\ ,\\
        -V_\tau(u)&\sim\sum_{I=1}^3\log\Gamma\left(-\frac{\pm u_i\pm u_j+\Delta_I}{\tau}\right)-\log\Gamma\left(-\frac{\pm u_i\pm u_j}{\tau}\right)\ ,
    \end{split}
\end{align}
where we suppressed the terms contributing to the exponent of the integrand of \eqref{eq:n=4sl3z} at $O(N)$.
Once again, splitting the dangerous Haar-measure-like factor, the matrix integral becomes
\begin{equation}
    \mathcal{I}(\Delta,\sigma,\tau)=\kappa^N\oint\prod_{i=1}^{N}du_i\prod_{\alpha\in \Delta^+_{G}}\left(1-e^{2\pi i\frac{\alpha_a(u)}{\tau}}\right)e^{-i\pi\mathbf{P}_+}e^{-V_\sigma(u)-\Tilde{V}_\tau(u)}\ .
\end{equation}
Then, under the constraint \eqref{eqn: two relevant branches of N=4 with other G} and
\begin{align}
    \Im\left(\frac{1-\tau+\Delta_I}{\sigma}\right)<0,\; \Im\left(\frac{\Delta_I}{\sigma}\right)>0,\;\Im\left(\frac{-\sigma+\Delta_I}{\tau}\right)>0,\; \Im\left(\frac{1+\Delta_I}{\tau}\right)<0\ ,
\end{align}
the second type of parallelogram ansatz \eqref{eqn: parallelogram ansatz (3)} is the extremum of the above integral with vanishing extremum value in the large $N$ limit. Finally, we obtain the large $N$ free energy of the index as follows:
\begin{equation}
    \log\mathcal{I}(\Delta_I,\sigma,\tau)\sim\begin{cases}
        -\frac{i\pi N(2N+1)\Delta_1\Delta_2\Delta_3}{\sigma\tau}\qquad&:Sp(N),SO(2N+1)\\
        -\frac{i\pi N(2N-1)\Delta_1\Delta_2\Delta_3}{\sigma\tau}\qquad&:SO(2N)
    \end{cases}\ .
\end{equation}
These are consistent with the free energy in the Cardy limit, where we take $\omega_{1,2} \to 0$ \cite{Amariti:2020jyx}. One finds that the above free energy is a half of that of the $\CN=4$ SYM with $SU(2N)$ gauge group \cite{Choi:2021rxi}, which is dual to AdS$_5 \times S^5$, at leading $O(N^2)$ order. This is expected from the dual AdS$_5$ gravity side since the 5d Newton constant is proportional to the volume of the internal space $\textrm{Vol} (\mathbb{RP}^5) = \textrm{Vol} (S^5/\mathbb{Z}_2) = \frac{1}{2}\textrm{Vol} (S^5)$.

\section{Discussion}\label{sec:discussion}

In this paper, we have established a universal large $N$ saddle point of the matrix model computing the superconformal index of 4d $\CN=1$ gauge theory. Our `parallelogram' ansatz gives $O(N^2)$ free energy, which accounts for the BPS black holes in AdS$_5$ for a holographic SCFT. We have shown that our saddle universally applies to a large class of gauge theories admitting a suitable large $N$ limit. It includes quiver gauge theories for $N$ D3-branes probing Calabi-Yau cone over Sasaki-Einstein 5-manifold, which is dual to type IIB superstring theory on AdS$_5 \times$ SE$_5$. 

We would like to emphasize that our result is not restricted to a particular limit or choice of fugacities and does not rely on ad hoc assumptions. We have shown that our ansatz is a genuine saddle point of the index integral (so that we can identify it with a solution to the equation of motion in the bulk, namely a black hole geometry) and works for a generic choice of parameters and theories. 
One of the new results is the large $N$ saddle point for the $SO(N)$, $Sp(N)$ gauge groups. Compared to that of the $SU(N)$ gauge group, only half of the parallelogram region is occupied by the eigenvalues. It will be interesting to find string theoretical interpretations of our large $N$ saddles.


Given the universality of the large $N$ superconformal index, it is tempting to say that whenever our saddle applies, the large $N$ gauge theory admits a holographic dual that includes a (BPS) black hole geometry. Quite generally, the large $N$ superconformal index has two competing saddles: one describing the confining phase with free energy scales as $O(1)$ and the other describing the deconfining phase with free energy scales as $O(N^2)$. This results in a confinement-deconfinement phase transition, which is dual to the Hawking-Page transition at a certain value of the chemical potential \cite{Choi:2018vbz}. 
This idea is further supported by noticing that our large $N$ saddle does not apply to `non-holographic' theories, such as SQCD in the Veneziano limit or the ones with `dense spectrum.' 

Unfortunately, our large $N$ saddle is `exceedingly general' to carve out non-holographic theories. In other words, our saddle may be applicable to SCFTs that do not have weakly-coupled gravity duals in AdS. For example, our saddle works for the $\CN=1$ gauge theories without any marginal coupling. In this case, we do not expect there to be tunable couplings (such as $\alpha'$) to suppress various higher-order corrections to supergravity. 
On the other hand, non-Lagrangian theories, such as the class $\mathcal{S}$ theories \cite{Gaiotto:2009we}, cannot be studied through our method since our analysis explicitly depends on the saddle point analysis of the gauge theory. We know that class $\mathcal{S}$ theories have a good holographic dual description in 11d supergravity \cite{Gaiotto:2009gz}. 

In general, we expect the CFTs to be holographically dual to a weakly-coupled AdS gravity if there is a gap in the higher-spin operator spectrum \cite{Heemskerk:2009pn} and the low-lying modes are sparse \cite{El-Showk:2011yvt}. The superconformal index captures many essential physics in the BPS sector, but it is not clear whether the index is sensitive enough to test these criteria to be holographic. 
It would be interesting to find a criterion that does not rely on a gauge theory description to decide whether the underlying theory is holographic or not, as was done in the case of 2d (S)CFTs \cite{Hartman:2014oaa, Benjamin:2015hsa, Benjamin:2015vkc}. 


Let us make a few remarks regarding possible future directions. In Section \ref{sec:more}, we have constructed the multi-cut solutions with equal filling fractions. However, in \cite{Choi:2021rxi}, it was shown that in some parameter region of the chemical potentials, the index of $\CN=4$ $U(N)$ SYM admits the multi-cut saddle points with unequal filling fractions. It would be interesting to find such solutions in $\CN=1$ gauge theories. Moreover, we would expect that there should exist many more large $N$ solutions to account for various black objects in the dual gravity side. One example is the hairy BPS black holes in AdS$_5$, which were recently studied in \cite{Choi:2023znd}.

Next, let us make a number of comments on the subleading corrections in $\frac{1}{N}$.
One interesting observation in our large $N$ analysis is that \eqref{eq:prefactor}, the prefactor $-i\pi \mathbf{P}_\pm$ after the $SL(3,\mathbb{Z})$ transformation, seems to give not only the leading $O(N^2)$ order but also the correct subleading corrections in $\frac{1}{N}$ of the full free energy of the index.
Such observation is based on the comparison with the Cardy formula \cite{Kim:2019yrz,Cabo-Bizet:2019osg,Cassani:2021fyv}, which is valid for arbitrary $N$ assuming no non-perturbative corrections in $\sigma, \tau$. Also note that for the $\mathcal{N}=4$ $SU(N)$ SYM at $\tau=\sigma$, it is verified through the Bethe-Ansatz approach at large $N$ in \cite{Mamroud:2022msu}.
Therefore, it is tempting to conjecture that the remaining integral of \eqref{eqn: SL(3,Z) transformed matrix integral-index} vanishes not only at the leading $O(N^2)$ order but also at subleading orders in $\frac{1}{N}$. In order to verify this conjecture, one needs to solve the large $N$ matrix model \eqref{eqn: SL(3,Z) transformed matrix integral-index} perturbatively in $\frac{1}{N}$, which is a challenging problem. However, for the first subleading order $O(N)$, we do not need to perturb the large $N$ solution itself. We can just insert the leading large $N$ solution to the effective action of $O(N)$ for the matrix model since the leading solution extremizes the effective action at $O(N^2)$. In fact, this is the strategy used in the four-derivative supergravity \cite{Bobev:2021qxx,Bobev:2022bjm,Cassani:2022lrk}. In our large $N$ matrix model, the obvious $O(N)$ terms of the effective action come from the fundamental matters and the Cartan parts of the rank-2 matters, which we neglected. These include the bifundamental matters between the gauge nodes $G_a$ and $G_b$ with $\textrm{rk}(G_a) = O(N),\; \textrm{rk}(G_b) = O(N^0)$. By evaluating their values at the parallelogram saddle point, it is straightforward to see that they indeed vanish at $O(N)$ since they are all given by the periodic integral in $\sigma$ or $\tau$ after the $SL(3,\mathbb{Z})$ transformation. However, there are other $O(N)$ contributions coming from the error of our large $N$ continuum approximation of the sum to the integral. In principle, it can be systematically analyzed, for instance, using the Euler-Maclaurin formula. In addition, there can also be contributions from the one-loop determinant in the large $N$ saddle point approximation, which is of $O(N\log N)$ at most. (For related study in the Bethe-Ansatz approach at large $N$, refer to \cite{Mamroud:2022msu}.) It would be interesting to study these subleading corrections and show that they indeed vanish at the parallelogram saddle point.
In the dual AdS$_5$ gravity side, it has been shown that the terms of $O(N)$ in \eqref{eq:prefactor} precisely match the four-derivative corrections to the on-shell action of the BPS black holes in minimal gauged supergravity \cite{Bobev:2021qxx,Bobev:2022bjm,Cassani:2022lrk}, thus supporting our conjecture at $O(N^1)$.

One may also want to study $O(\log N)$ correction to the free energy of the index.
Apart from the loop contributions in the large $N$ saddle point approximation, the $O(\log N)$ correction may occur when there are multiple large $N$ saddle points equally contributing to the index. It happens when the theory possesses a one-form symmetry \cite{ GonzalezLezcano:2020yeb, Amariti:2020jyx, Cassani:2021fyv}. Since our large $N$ saddle point of the gauge holonomies will spontaneously break such one-form symmetry, one can generate a set of solutions through the broken one-form symmetry action on that solution. They will equally contribute to the index, thus summing over their contributions yields $\log |Z|$ correction to the free energy of the index, where $|Z|$ is the order of the (abelian) one-form symmetry group $Z$. For example, in the gauge theories with only adjoint fields, such as the $\CN=4$ SYM and $\CN=1$ adjoint SQCDs, there indeed exists a one-form symmetry given by the center of the gauge group $Z(G)$, thus yielding the $\log |Z(G)|$ correction to the free energy. In addition, the Klebanov-Witten theory also carries a $\mathbb{Z}_N$ one-form symmetry, giving the $\log N$ correction to the free energy. In the dual AdS gravity side, the $O(\log N)$ correction is expected to be reproduced by one-loop supergravity around the black hole backgrounds \cite{Sen:2012kpz,Sen:2012cj,David:2021qaa}.

We would like to make a final remark about the periodicity of the index.
For simplicity, let us turn off all the flavor chemical potential $\delta^I$'s. Then, the unrefined index becomes
\begin{equation}\label{eq:unref2}
    \mathcal{I}(\omega)=\Tr \left[ (-1)^F e^{2\pi i\omega \left(J_+ + \frac{R}{2}\right)} \right]\ ,
\end{equation}
where we defined $\omega \equiv 2\delta = \sigma+\tau -1$ with $\sigma=\tau$. For a generic 4d $\CN=1$ SCFT, the superconformal $R$-charges of the operators are given mostly by irrational numbers. Therefore, the above index does not have any periodicity in general. One can view $\omega$ in this case as parameterizing the unclosed Lissajous curve on the maximal torus of the global symmetry.
However, for the theories with rational $R$-charged matters, such as the $\CN \geq 2$ SCFTs, \eqref{eq:unref2} will enjoy the periodicity.
Namely, the periodicity of the index \eqref{eq:unref2} sharply distinguishes the SCFTs with irrational $R$-charges and with rational $R$-charges.
This property will be particularly important when analyzing the supersymmetric spectral form factor studied in \cite{Choi:2022asl}.


\begin{acknowledgments}
We thank Minseok Cho, Seok Kim, Eunwoo Lee, Ki-Hong Lee, and Kimyeong Lee for the helpful discussions. 
This work is supported by a KIAS Individual Grant PG081602 at Korea Institute for Advanced Study (SC), and the National Research Foundation of Korea (NRF) Grant RS-2023-00208602 (SK, JS). 
The work of JS is also supported by the POSCO Science Fellowship of POSCO TJ Park Foundation.

\end{acknowledgments}


\bibliographystyle{JHEP}
\bibliography{ref}

\end{document}